\documentclass[english]{article}
\usepackage[T1]{fontenc}
\usepackage[latin9]{inputenc}
\usepackage{mathrsfs}
\usepackage{amsmath}
\usepackage{amssymb}
\usepackage{stmaryrd}
\usepackage{graphicx}
\usepackage{esint}
\usepackage{babel}
\begin{document}
\title{DYNAMICAL HORIZONS AND BLACK HOLE SOFT HAIR}
\author{Albert Huber\thanks{hubera@technikum-wien.at}, \thanks{albert.huber@uni-graz.at}}
\date{{\footnotesize UAS Technikum Wien, Höchstädtplatz 6, 1200 Vienna, Austria;
Karl-Franzens-University Graz, Schubertstraße 51, Graz, Austria}}
\maketitle
\begin{abstract}
In the present work, quasilocal Brown-York charges are derived that
coincide in the large sphere limit with the conserved supertranslation
hair and superrotation charges introduced by Hawking, Perry and Strominger
in \cite{hawking2016soft,hawking2017superrotation}. Given these charges,
a general scenario is outlined in which a non-rotating black hole
completely evaporates after its collapse due to particle creation
effects, whereby a genuine one-way traversable event horizon is never
formed, but merely a two-way traversable dynamical (resp. future trapping)
horizon. The formation of such a dynamical horizon has the consequence,
as is demonstrated, that quasilocal energy transported by the considered
charges, and thus information, can continuously escape through the
black hole horizon to infinity; a mechanism which, as is argued, could
possibly prevent information loss once the black hole formation and
evaporation process comes to an end.
\end{abstract}
{\footnotesize\textit{Key words: Black hole soft hair, dynamical horizons,
Hawking effect;}}{\footnotesize\par}

\section*{Introduction}

It is generally expected that black holes can evaporate due to contiuous
emission of thermal radiation. Responsible for this development are
particle creation effects first discovered by Hawking almost half
a century ago \cite{hawking1975particle}, whose existence has led
to one of the most persistent problems of modern theoretical physics:
the black hole information loss paradox.

This paradox occurs naturally when considering quantized matter collapsing
into a black hole that evaporates in finite time. More precisely,
when considering, a la Hawking, a quantized massless scalar field
in an approximately classical geometric background field in the context
of semiclassical Euclidean quantum gravity, this paradox arises naturally
from the fact that quantum correlations overcoming (by tunneling through)
the gravitational potential barrier of the black hole cause a continuous
emission of radiative modes to infinity. This has the consequence
that the asymptotic vacua at past and future null infinity do not
coincide, thereby implying that the past ground state of the field
must be a thermal state with respect to the future ground state and
vice versa. The latter is related to the fact that certain quantum
correlations are shielded by the black hole singularity and therefore
cannot be restored at future null infinity at the end of the evaporation
process. This, however, allows for the conclusion that the evolution
from low-entropy collapsing matter to high-entropy radiation must
result in a loss of quantum information.

Notwithstanding the fact that the validity of these results of Hawking's
work have not been questioned in any of the relevant literature on
the subject, this outcome of the process is generally deemed physically
implausible. After all, it would imply that black hole formation and
evaporation processes would inevitably lead to a breakdown of predictability
at the quantum level \cite{hawking1976breakdown,hawking1982unpredictability}.

In response to this obstacle, numerous approaches have been pursued
in the literature to solve the information loss problem and show that
quantum information is preserved in black hole evaporation processes;
see here e.g. \cite{hooft1996scattering,mathur2009information,mathur2009exactly,susskind1997black,susskind2003twenty,wald2001thermodynamics}
and references therein for an overview of the subject. 

Among the approaches mentioned, there is a fairly recent proposal
focusing on shock-wave-induced black hole soft hair, i.e. supertranslation
and/or superrotation charges conserved by asymptotic Bondi-Metzner-Sachs
(BMS) symmetries in asymptotically flat spacetimes. First considered
in \cite{flanagan2017conserved,hawking2016soft,hawking2017superrotation},
the existence of such conserved charges remarkably does not contradict
the no-hair theorem; for the simple reason that the supertranslated
black hole geometries considered are diffeomorphic to untranslated
ones and thus exhibit geometric properties that do not conflict with
the classic laws of black hole physics. In addition, the BMS invariance
of said charges suggests that there should be an infinite family of
degenerate vacua related by infinite-dimensional asymptotic symmetries
and thus an infinite number of conservation laws governing gravitational
scattering processes at null infinity. In light of these facts, the
authors of \cite{hawking2016soft,hawking2017superrotation} expressed
the idea that black hole soft hairs could encode the missing part
of the information that is lost during black hole formation and evaporation
process; thus possibly contributing to the solution of the information
paradox.

Yet, reservations have been expressed about this proposal \cite{bousso2017observable,bousso2017soft,mirbabayi2016dressed}
for two reasons: $i)$ soft hair should not be able to store information
and $ii)$ the arising quantum correlations should be too small to
purify Hawking radiation.

What is generally expected to be true, however, is that soft hair
leads to measurable effects such as the gravitational memory and spin
memory effects; as well as non-trivial quantum effects, where the
latter are likely to be of relevance to black hole physics \cite{haco2018black,pasterski2016new,strominger2016gravitational}.
Still, the majority of approaches set out in the literature still
treat soft hair exclusively classically without taking these very
quantum effects into account.

Against this background, this paper addresses two main objectives: 

The first one is to generalize the results of \cite{huber2022remark,huber2023quasilocal}
and show that, within the quasilocal Brown-York framework \cite{booth1999moving,brown2002action,brown1997energy,brown1993quasilocal},
integral expressions can be defined which asymptotically reduce -
as shown in section one and Appendix A of this work - to the conserved
supertranslation hair and superrotation charges derived in \cite{hawking2016soft,hawking2017superrotation}.
For this very purpose, a geometric framework is introduced that comprises
a spatially and temporally bounded spacetime with inner and outer
boundaries, where the inner boundary is formed by a dynamical black
hole horizon in the sense of Ashtekar et al. \cite{ashtekar2003dynamical,ashtekar2004isolated}.
Taking this framework as a starting point, a null geometric approach
is pursued, which serves as the basis for calculating the flux of
mass and/or radiant energy through said dynamical horizon in a non-stationary
black hole spacetime by performing a variation of the total gravitational
Hamiltonian of the theory (bulk plus boundary term). The latter then
leads to the aforementioned BMS charges and the associated black hole
soft hairs when the outer boundary of spacetime is shifted to infinity
in the large sphere limit.

The second objective, on the other hand, is to incorporate particle
creation effects into the formalism used. To this end, the classic
case of a quantized massless scalar field in a collapsing black hole
background is considered within the semiclassical approximation of
Euclidean quantum gravity. For said case, it is shown that boundary
charges can be defined that asymptotically reduce to BMS charges in
Bondi coordinates near null infinity; using in this context a semiclassical
(intergral law) version of the null Raychaudhuri equation derived
in section two of this work. Based on this result, the obtained expressions
are shown to allow for the consideration of quantum backreaction phenomena
caused by particle creation and the underlying Hawking effect; thus
describing, as is shown, 'quantum analogs' of conserved boundary charges
at the black hole horizon that naturally soften and take the form
of black hole supertranslation hairs as they approach past and/or
future null infinity in the large sphere limit. These quantum charges
are constructed using a semiclassical extension of the Bondi-Sachs
formalism briefly discussed in Appendix A of this paper.

Given these insights, a general scenario is outlined in the rear part
of the second section of the paper in which a black hole first forms
and then completely evaporates, whereby the event horizon - in the
spirit of earlier works by Hayward and Ashtekar \cite{ashtekar2020black,hayward2005disinformation}
- is replaced by a dynamical (resp. future trapping) horizon. Using
a model of a Vaidya-type sandwich spacetime presented in Appendix
B as an inspiration, it is then made clear that, according to the
scenario described, it should be possible for a flux of quasilocal
mass and/or radiant energy to escape through the dynamical horizon
of an evaporating black hole to future null infinity; where the quasilocal
charges responsible for this energy transfer are identified as those
derived in the paper, generalizing the supertranslation charges of
Hawking, Perry and Strominger. Eventually, by imposing conditions
on the mode functions of the scalar field, again derived with respect
to the toy model used, new arguments are provided for the conclusion
that particle creation effects in non-stationary black hole spacetimes
with locally defined dynamical horizons cannot cause information loss
at the quantum level.

Consequences of these discoveries are discussed towards the end of
the paper.

\section{Quasilocal Charges in Bounded and Unbounded Gravitational Fields }

In this section, using a null geometric approach to the quasilocal
Brown-York framework \cite{booth1999moving,brown2002action,brown1997energy,brown1993quasilocal,huber2023quasilocal},
an expression is derived for the mass and/or radiative energy flux
through a two-surface enclosing a nonstationary black hole spacetime.
To this avail, a system of integral expressions is defined that reduces
in an asymptically flat black hole background in the Bondi gauge close
to null infinity to the conserved supertranslation hair and superrotation
charges derived in \cite{flanagan2017conserved,hawking2017superrotation}.
Moreover, antipodal matching conditions, first introduced in \cite{hawking2017superrotation},
are discussed and implications of these conditions for the black hole
information loss paradox are pointed out. Eventually, following the
reasoning of \cite{huber2023quasilocal}, generalized results on mass
and radiative transfer in bounded and unbounded gravitational fields
are derived, which lead to quasilocal corrections to the Bondi mass-loss
formula and thus make clear that even more general boundary charges
can be constructed by the approach to be introduced.

\subsection{Geometric Framework}

In the following, a geometric framework similar to the one introduced
in \cite{huber2022remark,huber2023quasilocal} is considered, which
relies heavily on the Brown-York quasilocal formalism \cite{booth1999moving,booth2005horizon,brown2002action,brown1997energy,brown1993quasilocal}.
Specifically, a spatially and temporally bounded spacetime $(\mathcal{M},g)$
with manifold structure $\mathcal{M}=M\cup\partial M$ will be considered
with an exterior part $\partial M_{ext}$ and an an interior part
$\partial M_{int}$ such that $\partial M\equiv\partial M_{int}\cup\partial M_{ext}$.
The exterior part $\partial M_{ext}$ here is chosen in such a way
that $\partial M_{ext}\equiv\Sigma_{1}\cup\mathcal{B}\cup\Sigma_{2}$
applies, given that $\Sigma_{1}$ and $\Sigma_{2}$ are spatial boundary
parts, while $\mathcal{B}$ represents a timelike boundary portion.
By assumption, the considered spacetime is presumed to be foliated
by a collection of spacelike three-hypersurfaces $\{\Sigma_{t}\}$
that are given with respect to a future-directed time evolution vector
field $t^{a}=Nn^{a}+N^{a}$, where $N$ and $N^{a}$ are the corresponding
lapse function and shift vector field as usual and $n^{a}$ is the
normalized timelike generator leading to the the spacelike slicing
of $(\mathcal{M},g)$. Also, the timelike portion $\mathcal{B}$ of
$(\mathcal{M},g)$ is assumed to be foliated by a collection of two-surfaces
$\{\Omega_{t}\}$ such that $\mathcal{B}=\{\underset{t}{\cup}\Omega_{t}:\,t_{1}\leq t\leq t_{2}\}$.
At this same timelike boundary $\mathcal{B}$, the time-flow vector
field $t^{a}$ can then be assumed to take the form $t^{a}=\mathcal{N}v^{a}+\mathcal{N}^{a}$,
provided that $v^{a}$ is some timelike vector field tangent to $\mathcal{B}$
and orthogonal to $\Omega_{t}$ and $\mathcal{N}$ and $\mathcal{N}^{a}$
are the corresponding boundary lapse function and boundary shift vector
field, respectively. Furthermore, it will be assumed that there exists
an interior boundary $\partial M_{int}\equiv\mathcal{S}_{1}\cup\mathcal{H}\cup\mathcal{S}_{2}$,
where $\mathcal{H}$ is a hypersurface representing a (canonical)
dynamical horizon in the sense of Ashtekar et al. \cite{ashtekar2003dynamical,ashtekar2004isolated}.
That is to say, $\mathcal{H}$ is assumed to be a smooth, three-dimensional
submanifold (i.e. a spacelike or timelike hypersurface or in special
cases even null hypersurface) of spacetime that exhibits a foliation
$\{\mathcal{S}_{t}\}$ by marginally trapped surfaces such that relative
to each leaf of the foliation there exist two null normals $l^{a}$
and $k^{a}$ and two associated null expansion scalars $\Theta=q^{ab}\nabla_{a}l_{b}$
and $\Xi=q^{ab}\nabla_{a}k_{b}$, where $q^{ab}$ is the inverse of
the induced metric $q_{ab}=g_{ab}+l_{a}k_{b}+k_{a}l_{b}$, one of
which vanishes locally and the other of which is strictly negative,
i.e. $\Theta<0$ and $\Xi=0$ on $\mathcal{H}$. A dynamical horizon
is therefore nothing other than a special future trapping horizon
in the sense of \cite{hayward1994general}. Fig. 1 shows a schematic
three-dimensional representation of the spacetime manifold $M$ with
its boundaries. 

To further explain the notation, it may be noted that the the induced
three-metric at $\Sigma_{t}$ is $h_{ab}=g_{ab}+n_{a}n_{b}$. The
induced three-metric at $\mathcal{B}$, one the other hand, takes
the form $\gamma_{ab}=g_{ab}-u_{a}u_{b}$ , with $u^{a}$ denoting
a spatial unit vector field which is perpendicular to $\mathcal{B}$
and orthogonal to an associated temporal unit vector field $v^{a}$
tangent to $\mathcal{B}$. Moreover, in considering a further spacelike
vector field $s^{a}$ that is orthogonal to the timelike generator
$n^{a}$ of the spacelike foliation $\{\Sigma_{t}\}$, but generally
non-orthogonal to $v^{a}$ (in contrast to $u^{a}$, which is generally
non-orthogonal to $n^{a}$), one finds that the induced two-metric
$q_{ab}$ at $\Omega_{t}$ takes the form $q_{ab}=g_{ab}-u_{a}u_{b}+v_{a}v_{b}=g_{ab}+n_{a}n_{b}-s_{a}s_{b}$.
Using the latter relation in combination with the decompositions $n^{a}=\frac{1}{\sqrt{2}}(k^{a}+l^{a})$
and $s^{a}=\frac{1}{\sqrt{2}}(k^{a}-l^{a})$ of $n^{a}$ and $s^{a}$,
where $l^{a}$ and $k^{a}$ are null normals reducing locally to those
associated with a given leaf $\mathcal{S}_{t}$ of the foliation $\{\mathcal{S}_{t}\}$
of the dynamical horizon $\mathcal{T}$, one then finds that the induced
metric at said horizon takes the previously claimed form $q_{ab}=g_{ab}+l_{a}k_{b}+k_{a}l_{b}$.
Eventually, with respect to this induced metric, the boundary shift
vector reads $\mathcal{N}^{a}=q_{\;c}^{a}N^{c}$.

\begin{figure}
\includegraphics[viewport=-96.0222bp 0bp 400.413bp 547.726bp,scale=0.45]{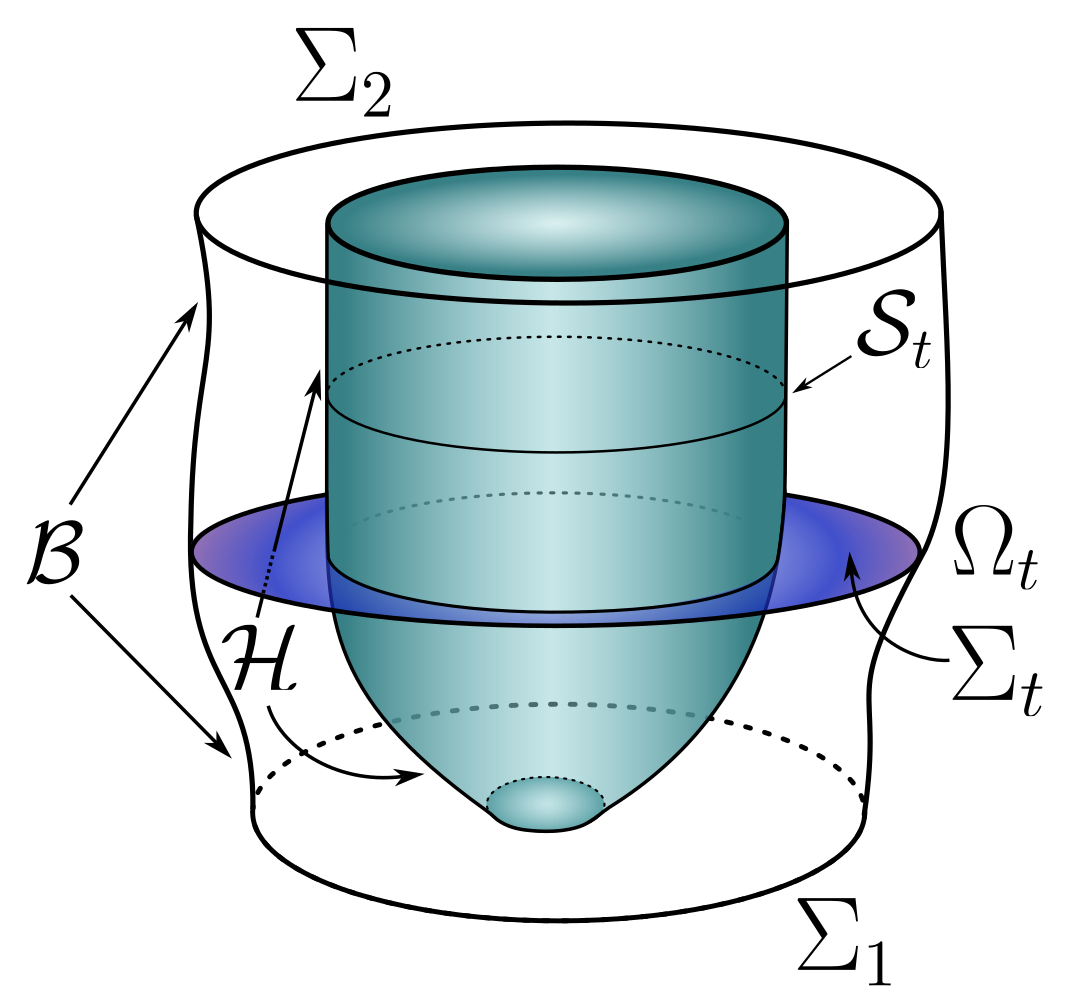}\caption{A schematic three-dimensional representation of the spacetime manifold
$M$ along with its boundaries.}
\end{figure}

\subsection{Quasilocal Brown-York Charges, Variation of the Gravitational Hamiltonian }

The starting point for all considerations to be made in this subsection
is the total gravitational Hamiltonian of the theory. Given a spacelike
hypersurface $\mathcal{C}$, this Hamiltonian is found to consist
of a bulk and a boundary part, meaning that $H=H_{Bulk}+H_{Boundary}$,
where $H_{Bulk}=\underset{\mathcal{C}}{\int}\mathcal{H}(N,\vec{N})d^{3}x$
and $H_{Boundary}=-\underset{\partial\mathcal{C}}{\oint}\mathfrak{h}(\mathcal{N},\vec{\mathcal{N}})d^{2}x$
are the corresponding bulk and boundary parts; thus giving rise to
the integral expression
\begin{equation}
H(N,\vec{N};\mathcal{N},\vec{\mathcal{N}})=\underset{\mathcal{C}}{\int}\mathscr{H}(N,\vec{N})d^{3}x-\underset{\partial\mathcal{C}}{\oint}\mathfrak{H}(\mathcal{N},\vec{\mathcal{N}})d^{2}x.
\end{equation}
Considering here the fact that the boundary of spacetime consists
of two individual parts, it is found that the Hamiltonian $(1)$ splits
in two different integral expressions from now on two be referred
as interior and exterior parts of the Hamiltonian, i.e.

\begin{align}
H_{0}^{Bulk}(N,\vec{N})= & \underset{\Sigma_{t}}{\int}\mathscr{H}(N,\vec{N})d^{3}x,\;H_{0}^{Boundary}(\mathcal{N},\vec{\mathcal{N}})=\underset{\Omega_{t}}{\int}\mathfrak{H}(\mathcal{N},\vec{\mathcal{N}})d^{2}x\\
H_{h}^{Bulk}(N,\vec{N})= & \underset{\mathcal{H}}{\int}\mathscr{H}(N,\vec{N})d^{3}x,\;H_{h}^{Boundary}(\mathcal{N},\vec{\mathcal{N}})=\underset{\mathcal{S}_{t}}{\int}\mathfrak{H}(\mathcal{N},\vec{\mathcal{N}})d^{2}x,\nonumber 
\end{align}
where $\mathscr{H}(N,\vec{N})$ represents the ADM Hamiltonian density
$\mathscr{H}=\frac{\sqrt{h}}{8\pi}(G_{ab}-8\pi T_{ab})t^{a}n^{b}=\frac{\sqrt{h}}{8\pi}(N\mathcal{H}+\mathcal{H}_{a}N^{a})$,
leading one to conclude that the bulk part $H_{Bulk}$ of the Hamiltonian
$(1)$ vanishes on-shell. The local Hamiltonian density $\mathfrak{H}(\mathcal{N},\vec{\mathcal{N}})$
at $\Omega_{t}$, on the other hand, can be written in the form $\mathfrak{H}=\frac{\sqrt{q}}{8\pi}\rho_{ab}t^{a}v^{b}=\frac{\sqrt{q}}{8\pi}(\mathcal{N}\mathfrak{h}+\mathfrak{h}_{a}\mathcal{N}^{a})$,
by virtue of the fact that the definitions $\mathfrak{h}=\frac{\sqrt{q}}{8\pi}\mathfrak{K}$
and $\mathfrak{h}_{a}=\frac{\sqrt{q}}{8\pi}u^{b}D_{a}v_{b}$ with
$\mathfrak{K}=q^{ab}\mathfrak{K}_{ab}=q^{ab}D_{a}u_{b}$ are used
in this context, where $v^{a}$ and $u^{a}$ are timelike and spacelike
vector fields tangential and orthogonal to $\mathcal{B}$, $\mathfrak{K}_{ab}$
is the extrinsic curvature calculated with respect to $u^{a}$ and
$D_{a}$ the covariant derviative at $\Omega_{t}$.

Sure enough, the form of the Hamiltonian depends on the choice of
the time evolution vector field $t^{a}$. In principle, any given
choice can reasonably be made for said vector field in absence of
exact and/or asymptotic symmetries of spacetime. Yet, as for later
purposes of this work, it will make sense to consider a null time-flow
vector field of the form $t^{a}=\sqrt{2}Nk^{a}$; a choice that will
only be legitimized in retrospect, in the further course of this section.
For now, let it just be noted that the main motivation for this particular
choice of $t^{a}$ comes from the theory of dynamical horizons, where
a very similar choice is made for the time-flow vector field, but
typically a specific ansatz for the lapse function of the geometry
is made. In what follows, however, there is no need for such a specific
choice of lapse, which is why no restrictions are placed on the latter.

Against this background, let it first be noted that the interior boundary
Hamiltonian $H_{h}^{Boundary}$ vanishes as $\Xi=0$ applies. This
can rather straightforwardly be concluded from the fact that, in view
of the special choice made fo $t^{a}$, the boundary Hamiltonian density
$\mathfrak{H}(\mathcal{N},\vec{\mathcal{N}})$ can be converted into
\begin{equation}
\mathcal{\mathfrak{H}}=\frac{\sqrt{2q}N\Xi}{8\pi};
\end{equation}
see here \cite{huber2023quasilocal} for further details. This form
for the Hamiltonian density, as should be noted, proves consistent
with the form of the quasilocal stress energy tensor calculated in
\cite{chandrasekaran2022brown} and \cite{jafari2019stress}, that
is, to the corresponding expressions for the energy densities (see
here equation $(3.44)$ on page $15$ of \cite{chandrasekaran2022brown}
and equation $(36)$ on page $16$ of \cite{jafari2019stress}, but
note that slightly different null geometric settings and notational
conventions are used).

To venture further, taking some other results of \cite{huber2022remark,huber2023quasilocal}
into account, let it be noted that the variation of the bulk part
of the exterior Hamiltonian yields an integral expression of the form

\begin{equation}
\mathfrak{L}_{t}H_{0}^{Bulk}=\underset{\Sigma_{t}}{\int}(\dot{N}\mathcal{H}+\mathcal{H}_{a}\dot{N}^{a}+\frac{N}{2}\mathcal{Q}_{ab}\dot{h}^{ab})\omega_{h}+\underset{\Omega_{t}}{\int}\Pi\omega_{q},
\end{equation}
where $\mathcal{Q}_{ab}=h_{a}^{\;c}h_{b}^{\;d}G_{cd}$ applies by
definition, $\mathfrak{L}_{t}$ denotes the Lie derivative with respect
to the time-flow vector field $t^{a}$ along $\Sigma_{t}$ and the
occurring boundary integral is defined with respect to an integrand
of the form $\Pi=\frac{N^{2}}{4\pi}(G_{ab}-8\pi T_{ab})k^{a}k^{b}$.
Accordingly, this variation of the Hamiltonian vanishes on-shell,
from which it can be concluded that, at the boundary of spacetime,
the gravitational and matter bulk-to-boundary inflow terms occurring
in $(4)$ cancel each other out. This, however, just implies the validity
of the full Einstein equations 
\begin{equation}
G_{ab}=8\pi T_{ab}
\end{equation}
 in the bulk and that of the associated scalar equation

\begin{equation}
G_{ab}k^{a}k^{b}=8\pi T_{ab}k^{a}k^{b}
\end{equation}
at the boundary of spacetime. Of course, an analogous calculation
can also be performed for the interior (horizon) part of the Hamiltonian
$H_{h}$.

Meanwhile, as far as the exterior part of the Hamiltonian is concerned,
a variation of the boundary part $H_{0}^{Boundary}$ yields a power
functional $\mathcal{P}_{0}$ of the form

\begin{align}
\mathcal{P}_{0}= & \mathcal{L}_{t}H_{0}^{Boundary}=\\
= & \frac{1}{4\pi}\underset{\Omega_{t}}{\int}\left[\dot{\mathcal{N}}\mathfrak{h}+\mathfrak{h}_{a}\dot{\mathcal{N}}^{a}+\frac{\mathcal{N}}{2}\mathfrak{Q}_{ab}\dot{q}^{ab}\right]\omega_{q},\nonumber 
\end{align}
given that $\mathcal{L}_{t}$ is the Lie derivative with respect to
the time-flow vector field $t^{a}$ along $\Omega_{t}$. This result,
as may be noted, is fully in line with known results from the literature
\cite{booth1999moving} and has previosuly been rederived in \cite{huber2022remark,huber2023quasilocal}.
For the specific choice of $t^{a}=Nk^{a}$, as shown in \cite{huber2023quasilocal},
one finds
\begin{align}
\mathcal{P}_{0}= & \mathcal{L}_{t}H_{0}^{Boundary}=\\
= & \frac{1}{4\pi}\underset{\Omega_{t}}{\int}\left[N\mathcal{L}_{k}N\Xi+N^{2}(\varkappa\Xi-\sigma'_{ab}\sigma'{}^{ab}+\omega'_{ab}\omega'{}^{ab}-8\pi T_{ab}k^{a}k^{b})\right]\omega_{q},\nonumber 
\end{align}
asssuming that $\varkappa=-(k\nabla)k_{a}l^{a}$ applies in this context.
In the large sphere limit, this expression reduces to the form

\begin{align}
\mathcal{P}_{0,\infty}=\underset{r\rightarrow\infty}{\lim}\mathcal{L}_{t}H_{0}^{Boundary} & =\\
=- & \frac{1}{4\pi}\underset{\mathcal{S}_{\infty}}{\int}\left[\vert\mathcal{\mathtt{N}}\vert^{2}+8\pi T_{kk}\right]d\Omega\nonumber 
\end{align}
in the asymptotically flat context when standard fall-off conditions
are assumed close to infinity (for further details, see here \cite{hayward1994general,penrose1984spinors}),
where, as may be noted, the definitions $T_{kk}=\underset{r\rightarrow\infty}{\lim}r^{2}T_{ab}k^{a}k^{b}$
and $\sigma'_{ab}\sigma'{}^{ab}=\vert\sigma'\vert^{2}=\frac{\vert\mathcal{\mathtt{N}}\vert^{2}}{r^{2}}$
have been used in this context. This makes it clear, however, that
the expression obtained coincides exactly with the Bondi mass-loss
formula discovered in \cite{bondi1962gravitational}, i.e. $\mathcal{P}_{0,\infty}\equiv\dot{M}_{B}$;
see here \cite{penrose1984spinors} page $426$ for comparison. As
may be noted, this is fully consistent with a classic result of \cite{brown1997energy}
(equation $(0.1)$ on page $3$) and with the results of \cite{jafari2019stress}
(equation $(57)$ on page $10$), the subject of which is the integral
law $\underset{\mathcal{S}_{\infty}}{\int}[\mathfrak{H}-\mathfrak{H}^{ref}]d^{2}x=\frac{1}{8\pi}\underset{\mathcal{S}_{\infty}}{\int}[\mathfrak{h}-\mathfrak{h}^{ref}]\omega_{q}=M_{B}$,
which occurs for the choice $\mathcal{N}^{a}=0$, $\mathcal{N}=N=1$,
where $M_{B}$ represents the Bondi mass and the reference term is
chosen as that of flat Minokwski space. Consequently, by performing
a variation of this expression with respect to the time-flow vector
field chosen, one finds (since the variation of the boundary reference
term is zero) integral law $(9)$ exactly reproduced.

Based on these findings, given two incoming and outgoing time-flow
vector fields of the form $t_{\pm}^{a}=\sqrt{2}Nk_{\pm}^{a}$, the
pair of quasilocal charges 
\begin{equation}
Q_{\pm}=\frac{1}{4\pi}\underset{\mathcal{I}^{\pm}}{\int}\left[\vert\mathcal{\mathtt{N}}_{\pm}\vert^{2}+8\pi T_{kk}^{\pm}\right]\varepsilon_{\pm q}
\end{equation}
can be specified by partial integration of $(9)$, where the definitions
$T_{kk}^{\pm}:=\underset{r\rightarrow\infty}{\lim}r^{2}T_{ab}k_{\pm}^{a}k_{\pm}^{b}$
and $\varepsilon_{+q}=du\wedge\omega_{q}$ resp. $\varepsilon_{-q}=dv\wedge\omega_{q}$
have been used with $u$ and $v$ denote the associated retarded and
advanced null time parameters, respectively. 

\subsection{Quasilocal Approach to Supertranslation Hair and Superrotation Charges,
Antipodal Matching Conditions}

In completely the same fashion as above, when making the choices $t_{+}^{a}=f\partial_{u}^{a}-\frac{1}{r}D^{A}f\partial_{A}^{a}+\frac{1}{2}D^{2}f\partial_{r}^{a}=fk_{+}^{a}+V_{+}^{a}$
with $\partial_{u}f=\partial_{r}f=0$ and $t_{-}^{a}=f\partial_{v}^{a}-\frac{1}{r}D^{A}f\partial_{A}^{a}+\frac{1}{2}D^{2}f\partial_{r}^{a}=fk_{-}^{a}+V_{-}^{a}$
with $\partial_{v}f=\partial_{r}f=0$, the quasilocal charges 

\begin{equation}
Q_{f}^{\pm}=\frac{1}{4\pi}\underset{\mathcal{I}^{\pm}}{\int}\left[\vert\mathcal{\mathtt{N}}_{\pm}\vert^{2}-\frac{1}{4}D_{a}D_{b}\mathtt{N}_{\pm}^{ab}+8\pi T_{kk}^{\pm}\right]f\varepsilon_{\pm q}
\end{equation}
can be deduced from $(8)$ by taking the large sphere limit in Bondi
coordinates close to null infinity, where some important details of
the calculation are given in Appendix $A$ of this work. To convince
oneself about the consistency of this result, one may take another
result of \cite{brown1997energy} into account (equation $(4.5)$
on page $9$), which shows that for a time-flow vector fields meeting
the condition $t_{+}^{a}=f\partial_{u}^{a}\vert_{\mathcal{I}^{+}}$
and $t_{-}^{a}=f\partial_{v}^{a}\vert_{\mathcal{I}^{-}}$ as well
$\mathcal{N}^{a}=0$ at the spatial boundary $\mathcal{B}$ of spacetime
it follows at leading order that
\begin{equation}
Q_{f}^{\pm}=\underset{r\rightarrow\infty}{\lim}H_{0}^{Boundary}=\frac{1}{4\pi}\underset{\mathcal{S}_{\infty}}{\int}fm_{B}^{\pm}\omega_{q},
\end{equation}
for $H_{0}^{Boundary}=\frac{1}{8\pi}\underset{\mathcal{S}_{\infty}}{\int}[\mathfrak{h}-\mathfrak{h}^{ref}]\omega_{q}$,
$f(x^{2},x^{3})=\underset{r\rightarrow\infty}{\lim}N(r,x^{2},x^{3})$,
given that $m_{B}(u,x^{2},x^{3})$ - resp. $m_{B}(v,x^{2},x^{3})$
- is the Bondi mass aspect of the geometry, thus giving an expression
that corresponds to Geroch's supermomentum \cite{brown1997energy,geroch1977asymptotic}
in case that the boundary reference term is chosen appropriately.
Yet, this allows one to conclude that
\begin{equation}
Q_{f}^{\pm}=\underset{r\rightarrow\infty}{\lim}H_{0}^{Boundary}=\frac{1}{4\pi}\underset{\mathcal{I}^{\pm}}{\int}f\dot{m}_{B}^{\pm}\varepsilon_{\pm q},
\end{equation}
giving an expression that immediately leads to $(11)$ by virtue of
the fact that $\partial_{u}m_{B}^{+}=\vert\mathcal{\mathtt{N}}_{+}\vert^{2}+\frac{1}{4}D_{a}D_{b}\mathtt{N}_{+}^{ab}+8\pi\mathfrak{T}_{uu}$
- resp. $\partial_{v}m_{B}^{-}=\vert\mathcal{\mathtt{N}}_{-}\vert^{2}+\frac{1}{4}D_{a}D_{b}\mathtt{N_{-}}^{ab}+8\pi\mathfrak{T}_{vv}$
- applies in Bondi coordinates close to null infinity if the fall-off
conditions presented in Appendix A of this work are taken into account.
That said, it becomes clear that the quasilocal charges derived are
manifestly identical with the soft supertranslation charges derived
in \cite{flanagan2017conserved,hawking2017superrotation}, thus making
it clear that the authors' results can be reproduced also within the
Brown-York quasilocal framework used. As may also be noted, the deduced
quasilocal charges also turn out to be identical in the asymptotic
regime to those constructed in \cite{donnay2019carrollian}, even
though a different quasilocal approach was adopted in the mentioned
work. 

For all these agreements, however, it is important to note that no
considerations regarding bulk symmetries or conserved charges on the
horizon were made. Only the asymptotic behavior of quasilocal Brown-York
charges resp. their time derivatives was investigated, without imposing
any specific constraints on the spacetime geometry. In that sense,
the results obtained prove to be both quite general and fully consistent
with previous results from the literature.

That said, taking next the fact into account that the momentum part
of the boundary Hamiltonian can be decomposed in the form $\mathfrak{h}_{a}=\Omega_{a}-D_{a}\ln\xi$,
where $\Omega_{a}=q_{\;a}^{c}K_{bc}s^{b}=-q_{a}^{\;b}k^{c}\nabla_{b}l_{c}$
is the so-called Háji$\hat{c}$ek rotation one-form \cite{booth2005horizon,gourgoulhon20063+},
one may choose $\mathcal{N}^{a}\equiv\Omega\varphi^{a}$ to be an
angular vector field, so that the boundary part of $(2)$ can be rewritten
in the form
\begin{equation}
\underset{\Omega_{t}}{\int}\mathfrak{h}_{a}\mathcal{N}^{a}\omega_{q}=\frac{1}{8\pi}\underset{\Omega_{t}}{\int}[\Omega\Omega_{a}\varphi^{a}+\Omega(\varphi D)\ln\xi]\omega_{q}=J_{\mathcal{S}_{t}}^{\Omega\varphi}+\frac{1}{8\pi}\underset{\Omega_{t}}{\int}\Omega(\varphi D)\ln\xi\omega_{q}.
\end{equation}
Taking then the large sphere limit and assuming that the relation 

\begin{equation}
\Omega\Omega_{a}\varphi^{a}=\mathcal{\mathtt{N}}_{a}Y^{a}
\end{equation}
admits a solution close to null infnity (with this being a clearly
justified assumption as $\Omega\varphi^{a}$ can in principle be chosen
arbitrarily), a suitable choice for the gauge term occurring in $(14)$
allows the further pair of charges 
\begin{equation}
Q_{Y}^{+}=\frac{1}{8\pi}\underset{\mathcal{I}_{-}^{+}}{\int}\mathcal{\mathtt{N}}_{a}Y^{a}\omega_{q},\;\frac{1}{8\pi}\underset{\mathcal{I}_{+}^{-}}{\int}\mathcal{\mathtt{N}}_{a}Y^{a}\omega_{q}=Q_{Y}^{-},
\end{equation}
to be defined, which coincide again exactly with the expressions given
in \cite{flanagan2017conserved,hawking2017superrotation}. Here, as
may be noted, the vector field $\mathcal{\mathtt{N}}_{a}$ is nothing
but the the angular momentum aspect of spacetime and $Y^{a}$ is a
conformal Killing vector field on the two-sphere given by $Y^{a}=D^{a}\chi+\varepsilon^{ab}D_{b}\kappa$
with $\chi$ and $\kappa$ representing solutions of the differential
relation $(D^{2}+2)\chi=(D^{2}+2)\kappa=0$ for $D^{2}:=D_{a}D^{a}.$ 

To summarize, the preceding arguments show that a variation of the
Brown-York-Hamiltonian $(2)$ yields on-shell expressions for supertranslation
charges $(11)$ and superrotation charges $(16)$ in null coordinates
in the Bondi gauge, both of which agree exactly with known results
from the literature.

As pointed out in \cite{hawking2017superrotation}, the validity of
the antipodal matching conditions

\begin{equation}
Q_{f,Y}^{+}=Q_{f,Y}^{-}
\end{equation}
ensures that the gravitational scattering problem is well-defined,
while additionally implying that an infinite number of charges are
conserved in relativsitic scattering processes. Such a condition therefore
constitutes a natural starting point for finding a possible resolution
to the information loss paradox.

As far as the latter is concerned, however, the situation is somewhat
complicated in that in stationary black hole spacetimes $\mathcal{I}^{-}$
happens to be a Cauchy hypersurface, whereas $\mathcal{I}^{+}$ fails
to be one. Still, the hypersurface $\mathcal{I}^{+}\cup\mathcal{H}^{+}$
constitutes a Cauchy surface, so that it makes sense - in the case
of massless fields - to infer that $Q_{f}^{-}=Q_{f}^{\mathcal{I}^{-}}$
and $Q_{f}^{+}=Q_{f}^{\mathcal{I}^{+}}+Q_{f}^{\mathcal{H}^{+}}$,
thus leading one to conclude that

\begin{equation}
Q_{f}^{\mathcal{H}^{+}}=Q_{f}^{\mathcal{I}^{-}}-Q_{f}^{\mathcal{I}^{+}}.
\end{equation}
The resultant charge $Q_{f}^{\mathcal{H}^{+}}$ represents a major
obstacle to solving Hawking's paradox, for its existence is linked
to an increase in entropy, and thus a loss of information, at the
quantum level.

The idea expressed in \cite{hawking2017superrotation} to overcome
this problem is now to consider a supertranslated black hole spacetime
resulting from 'throwing in' an asymmetric shock wave, and to explicitly
determine the form of the resultant charge $(18)$; a thought-provoking
approach that should clearly prove consistent with the results of
the present section. Notwithstanding, said approach will not be pursued
further below. 

Instead, the idea will be pursued of modelling - in full agreement
with the assumptions made in this section - the black hole horizon
as a dynamical horizon in the sense of Ashtekar et al. (resp. future
trapping in the sense of \cite{hayward1994general}). This changes
the entire setting, as now the black hole horizon is no longer modelled
for all times as a quasi-stationary hypersurface through which neither
matter nor radiation can escape. On the contrary, the horizon is now
modeled as a dynamical non-teleological geometric object that allows
Hawking radiation to escape to infinity. In combination with specific
fall-off conditions on the field modes of the scalar field, which
are derived in Appendix B by means of the model of a scalar field
in a Vaidya sandwich spacetime, this change of setup proves sufficient
to ensure that an information loss paradox in the sense of Hawking
should not occur. The main reason for this is that in the given setting
the black hole horizon is a dynamical horizon, which permits radiation
modes to escape and carry quasilocal energy, and hence information,
through the black hole horizon to infinity, thus allowing all information
encoded in the emitted radiation in principle to be recovered at null
infinity. Before going into details at this point, however, this section
will close by demonstrating that the results obtained so far can be
generalized in several respects by taking the results of \cite{huber2023quasilocal}
into account, the latter of which can be used to construct quasilocal
charges of a more general form than $(10)$ and $(11)$.

\subsection{Generalized Results: Mass and Radiation Transfer in Bounded and Unbounded
Gravitational Fields}

As a basis for constructing quasilocal charges that generalize those
depicted in $(10)$ and $(11)$, a generic ansatz of the form $t^{a}=\sqrt{2}Nk^{a}+V^{a}$
for the time evolution vector of spacetime will be made in the following,
where $V^{a}=-Ns^{a}+N^{a}=(O-N)s^{a}+\mathcal{N}^{a}$ has to be
satisfied for the sake of consistency. 

Knowing that a time-flow vector field for spacetimes without time-translation
symmetry can arguably almost arbitrarily be chosen, the question arises
as to which ansatz for the shift vector field $V^{a}$ should ideally
be made. One way to approach this question is to first consider a
black hole spacetime $(\mathcal{M},g)$ with time translation symmetry
and axial symmetry, according to which the time-flow vector field
can simply be chosen to be the Killing vector field of the geometry
(that is, a linear combination of temporal and angular Killing vector
fields). In such a geometric setting, it certainly makes sense to
make an ansatz of the form $V^{a}=\Omega\varphi^{a}$ close to the
Killing horizon of spacetime, where $\Omega$ and $\varphi^{a}$ constutite
the angular velocity of the black hole and the angular Killing vector
field of the geometry. Admittedly, the resulting quasilocal charge
can then also be meaningfully defined at null infinity. This, however,
should apparantly work out in the more general case of a dynamical
black hole spacetime, which is intriguing not least because, as shown
in \cite{huber2023quasilocal}, the ansatz made allows the derivation
of the Ashtekar-Krishnan version of the first law of black hole mechanics
(with respect to a part $\Delta\mathcal{H}$ of the dynamical horizon
$\mathcal{H}$) \cite{ashtekar2003dynamical}.

That said, given the above vector field, the boundary Hamiltonian
density takes the form

\begin{equation}
\mathcal{\mathfrak{H}}=\frac{\sqrt{q}}{8\pi}\left[\sqrt{2}N\Xi-\varGamma_{V}\right],
\end{equation}
by virtue that the definition $\varGamma_{V}=(K_{ab}-Kh_{ab})s^{a}V^{b}+\lambda(\mathcal{N}\mathcal{D})\eta$
is used in the present context. By a temporal variation of the corresponding
boundary term $H_{0}^{Boundary}$, one obtains from the above - taking
into account that $s^{a}=\frac{1}{\sqrt{2}}(k^{a}-l^{a})$ and $D_{a}s^{a}=q^{ab}\nabla_{a}s_{b}=\frac{1}{\sqrt{2}}(\Xi-\Theta)$
applies by definition - the generalized power functional 
\begin{align}
\mathscr{P}_{0}=\mathcal{L}_{t}H_{0}^{Boundary}= & \mathcal{P}_{0}+\mathcal{L}_{V}\underset{\Omega_{t}}{\int}\mathcal{\mathfrak{H}}\omega_{q}=\\
=\mathcal{P}_{0}+\sqrt{2}\underset{\Omega_{t}}{\int}(O-N)[\mathcal{L}_{k-l}\mathcal{\mathfrak{H}} & +(\Xi-\Theta)\mathcal{\mathfrak{H}}]\omega_{q}+\mathcal{L}_{\mathcal{N}}\underset{\Omega_{t}}{\int}\mathcal{\mathfrak{H}}\omega_{q}\nonumber 
\end{align}
where $\mathcal{P}_{0}$ is given by $(7)$. Yet, taking the fact
into account that an integral over the boundary of a boundary is zero,
one finds that
\begin{equation}
\mathcal{L}_{\mathcal{N}}\underset{\Omega_{t}}{\int}\mathcal{\mathfrak{H}}\omega_{q}=\underset{\Omega_{t}}{\int}D_{a}(\mathcal{\mathfrak{H}}\mathcal{N}^{a})\omega_{q}=0
\end{equation}
applies to the corresponding integral expression depicted in $(20)$.
Accordingly, one is thus led to onclude that relation $(20)$ can
be re-written in the form
\begin{equation}
\mathscr{P}_{0}=\mathcal{P}_{0}+\mathfrak{P}_{0},
\end{equation}
provided that the definition
\begin{equation}
\mathfrak{P}_{0}=\sqrt{2}\underset{\Omega_{t}}{\int}(O-N)[\mathcal{L}_{k-l}\mathcal{\mathfrak{H}}+(\Xi-\Theta)\mathcal{\mathfrak{H}}]\omega_{q}
\end{equation}
is used in the present context. By performing the large sphere limit,
then, relation $(22)$ can be further converted by taking $(9)$ into
account, thereby yielding the result
\begin{equation}
\mathscr{P}_{0,\infty}=\dot{M}_{B}+\mathfrak{P}_{0,\infty},
\end{equation}
where $M_{B}$ is the Bondi-Sachs mass. Consequently, one is lead
to conclude that the occurring expression $\mathfrak{P}_{0,\infty}=\underset{r\rightarrow\infty}{\lim}\mathfrak{P}_{0}$
- with $\mathfrak{P}_{0}$ being given by $(23)$ - gives rise to
quasilocal corrections to the Bondi mass loss formula, which, as first
shown in \cite{huber2023quasilocal}, can be expected to be zero if
and only if the function $O(x)$ and the boundary shift vector field
$\mathcal{N}^{a}(x)$ are chosen appropriately. In general, however,
the derived quasilocal corrections will be different from zero. Their
existence thus reveals that in generic geometric settings, there should,
according to the Brown-York quasilocal formalism, be exceptions from
the generally accepted rule: \textit{The mass of a system is constant
if and only if there is no news}.

The derived corrections should be taken into account, since by integrating
$(24)$ they allow the definition of the class of quasilocal charges,
which contains the class of charges given by $(11)$ as a special
case. Also, as may be noted, these corrections play a role in the
variation of the boundary part of the horizon Hamiltonian $H_{h}^{Boundary}$
(recall that the variation of the bulk part $H_{h}^{Bulk}$ vanishes
on-shell). A possible ansatz for the time-flow vector field leading
to such a horizon Hamiltonian is $t^{a}=\sqrt{2}Nk^{a}+V^{a}$ with
$V^{a}=\mathcal{O}s^{a}+2\Omega\varphi^{a}$ and $O=N+\mathcal{O}$
such that $\mathcal{O}=\frac{\varkappa}{4\pi(K-K_{ab}s^{a}s^{b})}$,
thereby making it clear that $\varkappa$ coincides with the surface
gravity of the black hole, where one has thus $-\varkappa=\gamma+\bar{\gamma}$
in spin-coefficients. The boundary part of the Hamiltonian then takes
the form

\begin{equation}
H_{h}^{Boundary}=\underset{\mathcal{S}_{t}}{\int}\mathfrak{H}d^{2}x=\underset{\mathcal{S}_{t}}{\int}[\sqrt{2}N\Xi+\frac{\varkappa}{8\pi}+2\Omega\Omega_{a}\varphi^{a}]\omega_{q}+\varDelta
\end{equation}
where the relations $\mathcal{N}\mathfrak{K}=Nk-(K_{ab}-Kh_{ab})N^{a}s^{b}-\lambda(\mathcal{N}\mathcal{D})\eta-(\mathcal{N}\mathcal{D})v_{a}u^{a}$
and $\mathfrak{h}_{a}\mathcal{N}^{a}=2\Omega[\Omega_{a}\varphi^{a}-(\phi D)\ln\xi]$
have been used to obtain this particular form of the result. 

As may be noted, the $\varDelta$-term occurring in $(25)$ is a gauge-dependent
term that can be eliminated by a suitable choice for $\xi(x)$. Thus,
in case that the latter is assumed to be so and, moreover, $(\mathcal{M},g)$
is assumed to exhibit an axisymmetric vacuum black hole four-geometry
that is invariant under time translations, where $\{\mathcal{S}_{t}\}$
shall be the folia of a non-expanding Killing horizon $\mathcal{H}^{+}$
that forms the interior boundary of spacetime and $\mathcal{E}$ denotes
the Komar energy evaluated at the horizon, equation $(25)$ can be
re-written in the form 
\begin{equation}
H_{h}^{Boundary}=\frac{\varkappa}{4\pi}\mathcal{A}+2\Omega\mathcal{J}=\mathcal{E},
\end{equation}
where it has been used that $\Xi=0$ as well as $\varkappa=const.$
and $\Omega=const.$ applies locally at the black hole horizon. Of
course, $\mathcal{A}$ and $\mathcal{J}$ denote here the area and
angular momentum of the black hole, respectively. 

Accordingly, it can be concluded that the Brown-York Hamiltonian agrees
with the Komar energy in the given case except for one eliminable
gauge term, thereby allowing Smarr's formula and the first law of
black hole mechanics to be reproduced (given that the black hole is
uncharged). In consequence, however, a variation (not necessarily
a temporal variation) of said boundary term yields the well-known
result 

\begin{equation}
\delta\mathcal{E}=\frac{\varkappa}{8\pi}\delta\mathcal{A}+\Omega\delta\mathcal{J},
\end{equation}
provided that $N\delta\Xi$ is locally zero at the sections of the
horizon (as would be the case for a temporal variation of $(26)$).

As may be noted, an analogous scenario occurs in the electrovac case
resp. even when the notion of the Killing Horizon is replaced by a
more general non-expanding null horizon (i.e. either a weakly isolated
or isolated horizon in the sense of \cite{ashtekar2004isolated}),
leading to the generalized first law of black hole mechanics derived
in \cite{ashtekar2000generic,ashtekar1999isolated}. In the latter
case, as explained in more detail in the upcoming section, it then
would have to be expected that corrections of the form $(24)$ occur
at infinity, at least provided that the weakly isolated or isolated
horizon results as a special case of a dynamical horizon. 

As described at the beginning of this section, such a dynamical horizon
constitutes the inner boundary of spacetime in the introduced geometric
setting and therefore forms a natural part of the presented model,
while in contrast no such horizon was considered in \cite{hawking2017superrotation}.
Yet, as will now be shown, the existence of such a horizon makes a
significant difference in modeling black hole formation and evaporation
processes. For according to such a scenario, no information loss paradox
occurs, since in the phases in which spacetime is nonstationary and
in which the black hole horizon is dynamical, radiation modes can
carry away quasilocal energy and thus information through the black
hole horizon to null infinity, thereby causing backreactions on the
black hole singularity until the latter vanishes completely. This
is to be explained in more detail below. 

\section{Brown-York charges and Black Hole Soft Hair: Classical and Quantum
fields }

The arguments presented so far have been entirely classical. This
is now to change. To this end, the framework introduced in the previous
section will be revisited below within the semiclassical framework
of Euclidean gravity. In doing so, it is shown that quasilocal charges
of the form $(10)$, $(11)$ and $(16)$ can be constructed analogously
in the semiclassical approximation to quantum gravity, including the
important special case of a quantized massless scalar field in a nonstationary
black hole spacetime. Furthermore, it is shown that the antipodal
matching conditions introduced in the previous section can be satisfied
if the black hole horizon is assumed to be given by a dynamical horizon
resp. a future trapping horizon in the sense of \cite{ashtekar2020black,hayward2005disinformation}.
To satisfy these matching conditions, constraints are imposed on the
mode function of the quantized massless scalar field, which are justified
in Appendix B using the simplified example of a scalar field in a
Vaidya background spacetime with a time-dependent monotonically decreasing
mass that becomes zero at the very end of the evaporation process.
Based on the results obtained, it is eventually concluded that particle
creation effects in nonstationary black hole spacetimes with dynamical
horizons should not cause information loss at the quantum level. 

\subsection{Nonstationary Black Holes, Particle Creation Effects and Information
Loss in a Nutshell}

To generalize the results obtained in the previous section and to
include quantum effects in the physical framework presented so far,
a brief overview of particle creation effects in curved spacetimes
is given in this subsection. To this end, the geometric setting considered
so far is modified in that a sandwich spacetime is now taken into
account, which describes a black hole formed by the collapse of a
massless scalar field that evaporates after a certain time. That is
to say, a spacetime $(M,g)$ is considered whose metric can be treated
essentially classically; at least as long as quantum backreaction
effects are not taken into account, which shall be assumed for the
time being. This spacetime contains three different regions: $i)$
an initial flat region, $ii)$ a dynamical curved black hole region,
and eventually again $iii)$ a final flat region. Concerning the considered
matter fields, the focus will be placed - similarly to Hawking's original
work \cite{hawking1975particle} on the subject - on the case of a
massless, quantized scalar field $\varphi(x)$, which constitutes
a solution of the covariant wave equation

\begin{equation}
\boxempty\varphi=\nabla_{a}\nabla^{a}\varphi=0.
\end{equation}
As for any field in a generic curved geometric background, solutions
of said equation cannot be decomposed into its positive and negative
frequency parts, since positive and negative frequencies have no invariant
meaning in a curved spacetime when the latter lacks time translation
symmetry. Yet, since the spacetime model considered contains an initial
flat region, one can find the decomposition
\begin{equation}
\varphi=\overset{\infty}{\underset{k=1}{\sum}}[f_{k}^{-}a_{k}^{-\dagger}+\bar{f}_{k}^{-}a_{k}^{-}]
\end{equation}
of $\varphi(x)$ such that the purely ingoing mode functions $\{f_{k}^{-}(x)\}$
and $\{f_{k}^{-\ast}(x)\}$ contain only positive frequencies with
respect to the initial retarded Minkowskian time coordinate. Here,
$a_{k}^{-}$ and $a_{k}^{-\dagger}$ are latter operators satisfying
the standard canonical commutation relations for massless bosonic
fields. 

The above decomposition of $\varphi(x)$ is of course not unique.
Still, a decomposition of $\varphi(x)$ with respect to the final
retarded Minkowskian time coordinate and associated purely outgoing
field modes $\{f_{k}^{+}(x)\}$ and $\{f_{k}^{+\ast}(x)\}$ generally
cannot be performed due to the fact that $\mathcal{I}^{+}$, quite
in contrast to $\mathcal{I}^{-}$, fails to be a Cauchy hypersurface.
One way out of this dilemma is to carry out the decomposition of $\varphi(x)$
with respect to $\mathcal{I}^{+}\cup\mathcal{H}$, where $\mathcal{H}$
is the black hole horizon. The reason is that $\mathcal{I}^{+}\cup\mathcal{H}$
is actually a Cauchy horizon. Given this particular choice, the decomposition
reads

\begin{equation}
\varphi=\overset{\infty}{\underset{k=1}{\sum}}[f_{k}^{+}a_{k}^{+\dagger}+\bar{f}_{k}^{+}a_{k}^{+}+g_{k}b_{k}^{\dagger}+\bar{g}_{k}b_{k}],
\end{equation}
where $\{g_{k}(x)\}$ and $\{g_{k}^{\ast}(x)\}$ are horizon modes
and $a_{k}^{+\dagger}$, $a_{k}^{+}$ and $b_{k}^{\dagger}$, $b_{k}$
are creation and annihilation operators satisying the canonical commutation
relations
\begin{equation}
[b_{k},b_{l}]=[b_{k},a_{l}^{\pm}]=[b_{k},a_{l}^{\pm\dagger}]=[a_{k}^{\pm},a_{l}^{\pm}]=0,\;[a_{k}^{\pm},a_{l}^{\pm\dagger}]=[b_{k},b_{l}^{\dagger}]=\delta_{kl}.
\end{equation}
Given that spacetime is flat at early and late times, it can thus
be concluded that the pairs of annihilation operators $a_{k}^{\pm}$
give rise to initial and final asymptotic vacuum states $\vert0\rangle_{\pm}$,
which both states are solutions of the equations $a_{k}^{\pm}\vert0\rangle_{\pm}=0$.
Since the condition of asymptotic completeness \cite{haag2012local,hawking1982unpredictability}
is however clearly not fulfilled under the given geometric circumstances
(as the model considered lacks a unique Poincaré invariant vacuum
state), these asymptotic vacuum states must be inherently different
from each other. Also the form of the final mode functions $\{f_{k}^{+}(x)\}$
will differ from that of the initial mode functions $\{f_{k}^{-}(x)\}$.
Yet, since the latter form a complete basis for solutions of the wave
equation $(28)$ and massless fields are completely determined by
their initial data at past null infinity, the mode functions and ladder
operators in the future asymptotic region are related with those in
the past asymptotic region by the Bogoliubov transformations

\begin{equation}
f_{k}^{+}=\overset{\infty}{\underset{m=1}{\sum}}(\alpha_{km}f_{m}^{-}+\beta_{km}\bar{f}_{m}^{-}),\;g_{k}=\overset{\infty}{\underset{m=1}{\sum}}(\gamma_{km}f_{m}^{-}+\eta_{km}\bar{f}_{m}^{-})
\end{equation}
and

\begin{equation}
a_{k}^{+}=\overset{\infty}{\underset{m=1}{\sum}}(\alpha_{km}^{*}a_{m}^{-}-\beta_{km}^{*}a_{m}^{-\dagger}),\;b_{k}=\overset{\infty}{\underset{m=1}{\sum}}(\gamma_{km}^{*}a_{m}^{-}-\eta_{km}^{*}a_{m}^{-\dagger}),
\end{equation}
where the onditions $\overset{\infty}{\underset{m=1}{\sum}}(\alpha_{km}\alpha_{mj}^{*}-\beta_{km}\beta_{mj}^{*})=\delta_{kj}$
and $\overset{\infty}{\underset{m=1}{\sum}}(\alpha_{km}\beta_{mj}-\beta_{km}\alpha_{mj})=0$
as well as $\overset{\infty}{\underset{m=1}{\sum}}(\gamma_{km}\gamma_{mj}^{*}-\eta_{km}\eta_{mj}^{*})=\delta_{kj}$
and $\overset{\infty}{\underset{m=1}{\sum}}(\gamma_{km}\eta_{mj}-\eta_{km}\gamma_{mj})=0$
need to be satisfied for the sake of consistency. Solving the relation
$a_{k}^{+}\vert0\rangle_{+}=\overset{\infty}{\underset{m=1}{\sum}}(\alpha_{km}^{*}a_{m}^{-}-\beta_{km}^{*}a_{m}^{-\dagger})\vert0\rangle_{+}=0$
then yields
\begin{equation}
\vert0\rangle_{+}=Nexp(-\frac{1}{2}\overset{\infty}{\underset{m,n=1}{\sum}}\chi_{mn}a_{m}^{-\dagger}a_{n}^{-\dagger})\vert0\rangle_{-}
\end{equation}
for some normalization factor $N$ with $\chi=\frac{1}{2}(\alpha^{-1}\beta+(\alpha^{-1}\beta)^{T})$,
thereby showing that the asymptotic vacuum ground states of the field
do not coincide. 

Based on the fact that a mode of positive frequency at future null
infinity, at late times, matches onto mixed modes of positive and
negative frequencies at past null infinity, one thus comes to the
conclusion that the Bogoluibov coefficients in $(32)$ and $(33)$
are given in such a way that $\underset{m}{\sum}\beta_{km}\beta_{mj}^{*}=e^{-\pi\kappa^{-1}(\omega_{k}+\omega_{j})}\underset{m}{\sum}\alpha_{km}\alpha_{mj}^{*}$
and $(\beta\beta^{*})_{ii}\propto(e^{2\pi\kappa^{-1}\omega_{i}}-1)^{-1}$.
Accordingly, based on the fact that the corresponding coefficient
matrices $\beta_{km}$ and $\beta_{km}^{*}$ can be determined to
be directly proportional to the invertible matrices $\alpha_{km}$
and $\alpha_{km}^{*}$ in this particular case, one finds \cite{hawking1975particle}

\begin{equation}
N_{k}={}_{-}\langle0\vert a_{k}^{+\dagger}a_{k}^{+}\vert0\rangle_{-}=\overset{\infty}{\underset{l=1}{\sum}}\vert\beta_{kl}\vert^{2}=\overset{\infty}{\underset{l=1}{\sum}}\frac{\Gamma_{kl}}{e^{\frac{2\pi\omega_{k}}{\kappa}}-1}.
\end{equation}
The semiclassical arguments used thus clearly indicate that the final
quantum state of the field should be a mixed thermal state with Planck
spectrum of the form $(35)$ after the black hole has evaporated completely,
where $\kappa$ is the surface gravity of the black hole. One therefore
is led to conclude that that a given pure initial quantum state $\rho_{in}\equiv\vert\Psi\rangle\langle\Psi\vert$,
where $\vert\Psi\rangle$ is some Fock state constructed from $\vert0\rangle_{-}$,
inevitably evolves into a mixed state $\rho_{out}$ in the course
of black hole evaporation. The latter leads to an increase in entropy
and thus to a loss of quantum information; something that seems to
conflict with the standard rules of quantum field theory in flat Minkowski
space. 

That said, the exact way in which this loss of quantum information
manifests itself with respect to the quasilocal charges constructed
in the previous section will now be eludicated in the upcoming subsection
of this work, where charges of the form $(10)$, $(11)$ and $(16)$
are constructed in the framework of semiclassical Euclidean quantum
gravity. In doing so, as will be shown, an essential step is to take
into account those quantum backreaction effects that have remained
unconsidered in the present subsection.

\subsection{Boundary Charges and Quantum Backreaction}

So far, quantum backreaction effects of matter on the geometry of
spacetime have been neglected. In the short term, this may well be
justified for black holes in quasi-stationary equilibrium. Since the
continuous emission of Hawking radiation due to particle creation
effects clearly leads to a change in the geometry of the black hole
over time, however, these effects cannot simply be ignored in the
long run. On the contrary, these quantum backreaction effects must
clearly be taken into account in the definition of quasilocal Brown-York
charges; now to be generalized in the semiclassical framework of Euclidean
quantum gravity.

As a basis for the latter, it may be noted that the stress-energy
operator corresponding to a quantized massless scalar field $\varphi(x)$
of the form $(28)$ reads

\begin{equation}
T_{ab}=\nabla_{a}\varphi\nabla_{b}\varphi,
\end{equation}
meaning that one actually considers a distributional bitensor field
$T_{ab}(x,x')$ taking values at two different spacetime points $x$
and $x'$, which results from applying a bidifferential operator on
the bilinear product of operator-valued distributions (the field operators).
To obtain a well-defined tensor distribution from this expression
that is solely defined at a single point, it has proven convenient
to calculate the renormalized expectation value $\langle T_{ab}\rangle_{ren}(x)\equiv\langle T_{ab}\rangle_{\omega,ren}(x)$
of $T_{ab}(x,x')$ in some (algebraic) state $\omega$, where the
same state can be used as a basis for a GNS construction of the corresponding
Hilbert space. The calculation of $\langle T_{ab}\rangle_{ren}(x)$,
however, usually proves to be a subtle, technically involved issue,
which is essentially due to the fact that the point limit $x\rightarrow x'$
of $T_{ab}(x,x')$ is singular. Notwithstanding this, said expectation
value can be calculated by different techniques known from the literature
(see here, for example, \cite{birrell1984quantum,fulling1989aspects,mukhanov2007introduction,wald1994quantum}
and references therein for further information), which allow the divergent
parts of $\langle T_{ab}\rangle(x)$ to be removed, thereby giving
a suitably regularized version of this expectation value a mathematically
well defined meaning. Whereas some of these techniques focus primarily
on the renormalization of coupling constants in the gravitational
action, others focus more on controlling the short distance behavior
of the two-point distribution $\langle\varphi(x)\varphi(x')\rangle$
in order to calculate $\langle T_{ab}\rangle_{ren}(x)$ from $T_{ab}(x,x')$
in the point limit $x\rightarrow x'$. These latter so-called point-splitting
methods \cite{christensen1978regularization,christensen1976vacuum,dewitt1975quantum,fulling1989aspects,fulling1978singularity,wald1994quantum}
additionally reveal the fact that $\langle T_{ab}\rangle_{ren}(x)$
can only be reasonably calculated in a suitable class of states, which,
as is generally agreed, is the class of (quasi-free) Hadamard states.
These states not only allow the expectation value of the stress-energy
tensor to be calculated in rigorous fashion (e.g. by using the Hadamard
renormalization approach), but also show - by Wald's theorem \cite{wald1977back,wald1994quantum}
- that $\langle T_{ab}\rangle_{ren}(x)$ is only uniquely defined
up to the ambiguity of adding local curvature terms.

This all plays a role for the generalization of the quasilocal Brown-York
charges derived in the previous section. The reason for this is that
knowledge of the expectation value $\langle T_{ab}\rangle_{ren}$
is crucial for the description of quantum backreaction effects on
the geometry of spacetime in the context of the semiclassical Euclidean
quantum gravity. This, in turn, is due to the fact that the central
role in such a description is played by the semiclassical Einstein
equations

\begin{equation}
G_{\;b}^{a}=8\pi\mathcal{T}_{\;b}^{a}=8\pi\langle T_{\;b}^{a}\rangle_{ren}-\varTheta_{\;b}^{a},
\end{equation}
which differ by higher order corrections from the classical Einstein
equations $(5)$. These higher order corrections are all encoded in
the object $\mathcal{T}{}_{ab}=\langle T_{ab}\rangle_{ren}-\frac{1}{8\pi}\varTheta_{ab}$;
respectively, to be more precise, in the object
\begin{equation}
\varTheta_{\;b}^{a}=c_{1}H_{1\;b}^{a}+c_{2}H_{2\;b}^{a}+c_{3}H_{3\;b}^{a},
\end{equation}
which is given with respect to the renormalization constants $c_{1}$,
$c_{2}$ and $c_{3}$ and the tensor fields $H_{1\;b}^{a}$, $H_{2\;b}^{a}$
and $H_{3\;b}^{a}$ 
\begin{align}
H_{1\;b}^{a} & =2\nabla^{a}\nabla_{b}R-2\boxempty R\delta_{\;b}^{a}+2RR_{\;b}^{a}-\frac{1}{2}R^{2}\delta_{\;b}^{a},\\
H_{2\;b}^{a} & =2\nabla_{b}\nabla^{c}R_{\;c}^{a}-\boxempty G_{\;b}^{a}+2R_{\;c}^{a}R_{\;b}^{c}-\frac{1}{2}\delta_{\;b}^{a}R_{\;d}^{c}R_{\;c}^{d},\nonumber \\
H_{\epsilon3\;b}^{a} & =-\frac{1}{2}\delta_{\;b}^{a}R_{cdef}R^{cdef}+2R_{bcde}R^{acde}-4\boxempty R_{\;b}^{a}+2\nabla^{a}\nabla_{b}R\nonumber \\
 & -4R_{\;c}^{a}R_{\;b}^{c}+4R^{cd}R_{\;cbd}^{a};\nonumber 
\end{align}
see here, for instance, \cite{birrell1984quantum} for further information.

To set up $(37)$, it must therefore be ensured that $\langle T_{ab}\rangle_{ren}(x)$
and thus $\mathcal{T}_{\;b}^{a}$ can actually be calculated. If the
latter is the case, the obvious idea of how to transition from the
classical Brown-York charges derived in the previous section to the
corresponding quantum charges is to make the replacement $T_{\;b}^{a}\rightarrow\mathcal{T}_{\;b}^{a}$
in relations $(4-10)$. This yields, among other things, the result

\begin{align}
\mathcal{L}_{t} & H_{h}^{Boundary}=\\
=\frac{1}{4\pi}\underset{\mathcal{S}_{t}}{\int} & \left[N\mathcal{L}_{k}N\Xi+N^{2}[\varkappa\Xi-\sigma'_{ab}\sigma'{}^{ab}+\omega'_{ab}\omega'{}^{ab}-8\pi\mathcal{T}{}_{ab}k^{a}k^{b}]\right]\omega_{q},\nonumber 
\end{align}
with $\mathcal{S}_{t}\subset\mathcal{H}$ and thus 

\begin{align}
\mathcal{L}_{t}H_{h}^{Boundary} & =\\
=- & \frac{1}{4\pi}\underset{\mathcal{S}_{\infty}}{\int}\left[\vert\mathcal{\mathtt{N}}\vert^{2}+8\pi\mathcal{T}{}_{kk})\right]d\Omega,\nonumber 
\end{align}
in the large sphere limit, where the definition $\mathcal{T}{}_{kk}:=\underset{r\rightarrow\infty}{\lim}[r^{2}(\langle T_{ab}\rangle_{ren}k^{a}k^{b}-\varTheta_{ab}k^{a}k^{b})]$
has been used. 

So as to be able to consider said charges also within the semiclassical
regime of Euclidean quantum gravity for the given geometric model
of a collapsing and evaporating black hole spacetime, it shall be
assumed in the following that the renormalized expectation value of
the stress-energy tensor can actually be calculated at early and late
times (i.e. near past and future null infinity) from the asymptotic
vacuum states $\vert0\rangle_{\pm}$. That this is indeed possible
is by no means obvious, as the geometric model considered is dynamical
and thus lacks temporal and spatial translation symmetry as well as
a preferred Poincaré invariant vacuum state. Notwithstanding that,
it may however be possible to single out within the flat asymptotic
regions of spacetime the preferred asymptotic vacuum states $\vert0\rangle_{\pm}$
as well as associated preferred Hadamard states $\omega^{\pm}$ relative
to which the pairs $\langle T_{ab}^{\pm}\rangle_{ren}:=\langle T_{ab}\rangle_{\omega^{\pm},ren}$
and $\mathcal{T}_{ab}^{\pm}:=\langle T_{ab}^{\pm}\rangle_{ren}-\frac{1}{8\pi}\varTheta_{ab}$
of tensor distributions can be calculated (whereby, of course, one
and the same renormalization scheme must be used). 

Although the assumptions made are, as can be seen, quite strong, it
should be noted that they are both legitimate and in some ways even
substantial. For without a corresponding selection of asymptotic states
with a reasonably decent physical behavior, it would be extremely
hard to even posit Hawking's information paradox, let alone discuss
quantum backreaction effects and their physical consequences for the
evolution and ultimate fate of an evaporating black hole spacetime.
The assertions made therefore seem justified at least to the extent
that they allow particle creation effects and any alleged loss of
quantum information resulting from them to be discussed.

That said, it is worth investigating the integral expression $\underset{r\rightarrow\infty}{\lim}H_{h}^{Boundary}=\int\mathcal{L}_{t}H_{h}^{Boundary}du$
(resp. $\underset{r\rightarrow\infty}{\lim}H_{h}^{Boundary}=\int\mathcal{L}_{t}H_{h}^{Boundary}dv$)
a little closer, which occurs naturally in case that $t^{a}=\sqrt{2}Nk_{+}^{a}$
(resp. $t^{a}=\sqrt{2}Nk_{-}^{a}$) is chosen as the null time-flow
vector field of the geometry. This expression contains an integral
term of the form 
\begin{equation}
\underset{\mathcal{I}^{\pm}}{\int}N^{2}\dot{\Xi}\varepsilon_{\pm q}=\underset{\mathcal{I}^{\pm}}{\int}N^{2}[\kappa\Xi-\Xi^{2}-\sigma'_{ab}\sigma'{}^{ab}+\omega'_{ab}\omega'{}^{ab}-8\pi G{}_{k_{\pm}k_{\pm}}]\varepsilon_{\pm q},
\end{equation}
provided that the definition $G_{k_{\pm}k_{\pm}}:=\underset{r\rightarrow\infty}{\lim}r^{2}G_{ab}k_{\pm}^{a}k_{\pm}^{b}$
is used in the present context. Using $(37)$ to set up the relation
$G_{k_{\pm}k_{\pm}}=8\pi\mathcal{T}^{\pm}{}_{k_{\pm}k_{\pm}}$, where
$\mathcal{T}^{\pm}{}_{k_{\pm}k_{\pm}}:=\underset{r\rightarrow\infty}{\lim}[r^{2}(T_{ab}^{\pm}\rangle_{ren}k_{\pm}^{a}k_{\pm}^{b}-\frac{1}{8\pi}\varTheta_{ab}k_{\pm}^{a}k_{\pm}^{b}]$
applies by definition, one may argue that the resulting integral relation
\begin{equation}
\underset{\mathcal{I}^{\pm}}{\int}N^{2}\dot{\Xi}\varepsilon_{\pm q}=\underset{\mathcal{I}^{\pm}}{\int}N^{2}[\kappa\Xi-\Xi^{2}-\sigma'_{ab}\sigma'{}^{ab}+\omega'_{ab}\omega'{}^{ab}-8\pi\mathcal{T}^{\pm}{}_{k_{\pm}k_{\pm}}]\varepsilon_{\pm q},
\end{equation}
constittes a semiclassical version of the null Raychaudhuri equation.
This can be understood as follows: 

The final term of the obtained relation in the form includes integral
expressions of the form $\int\mathcal{T}^{+}{}_{k_{+}k_{+}}du$ and
$\int\mathcal{T}^{-}{}_{k_{-}k_{-}}dv$, which may be viewed as subject
to the averaged null energy condition (ANEC)

\begin{equation}
\int\langle T^{+}\rangle_{ren}{}_{k_{+}k_{+}}du\geq0,\;\int\langle T^{-}\rangle_{ren}{}_{k_{-}k_{-}}dv\geq0.
\end{equation}
This condition states that along a complete null curve, any negative
energy fluctuations of a quantum field must be balanced by positive
energy fluctuations. Matter that violates the ANEC could, if it existed,
be used to build time machines \cite{friedman1993topological,friedman1995topological,morris1988wormholes}
and violate the second law of thermodynamics \cite{wall2010proving}.
In contrast to most other energy conditions discussed in relativity
(dominant, strong, weak, zero, etc.), however, there are no known
counterexamples of the ANEC in quantum field theories (assuming that
the null geodesic is achronal).

That said, given that the ANEC $(44)$ proves valid, it can be expected
that also the conditions
\begin{equation}
\underset{\mathcal{I}^{\pm}}{\int}\mathcal{T}^{\pm}{}_{k_{\pm}k_{\pm}}\varepsilon_{\pm q}\geq0
\end{equation}
are met. The latter can be concluded if the following two mild assumptions,
which will be made below, prove to be correct: $i)$ The additional
angular integration of $(44)$ still to be carried out has no influence
on the positivity of the considered integral terms, i.e. $\underset{\mathcal{I}^{\pm}}{\int}\langle T^{\pm}\rangle_{ren}{}_{k_{\pm}k_{\pm}}\varepsilon_{\pm q}\geq0$.
$ii)$ The higher-curvature corrections become small enough close
to past and future null infinty in the sense that $\underset{\mathcal{I}^{\pm}}{\int}\langle T^{\pm}\rangle_{ren}{}_{k_{\pm}k_{\pm}}\varepsilon_{\pm q}\geq\frac{1}{8\pi}\underset{\mathcal{I}^{\pm}}{\int}\varTheta{}_{k_{\pm}k_{\pm}}\varepsilon_{\pm q}$
is satisfied. By these two assumptions, it then becomes clear that
the generalized ANEC $(45)$ is met. The validity of this relation
guarantees the well-definedness of the left hand sides of $(42)$
and $(43)$ even in the event that $\mathcal{T}^{\pm}{}_{k_{\pm}k_{\pm}}<0$
applies and null rays may not be focussing. This, however, suggests
that the resuting integral law can be viewed as a semiclassical generalization
of the null Raychadhuri equation. 

To derive this result, even more assumptions have been made (on top
of those already made before). However, it may be noted said assumptions
can again be justified to a reasonable extent on the basis of the
following two observations: On the one hand, all higher-curvature
corrections resulting from $\varTheta{}_{k_{\pm}k_{\pm}}$ should
be of higher order and therefore actually be very small close to future
null infinity. On the other hand, all reasonable matter distributions
are expected to meet the ANEC $(45)$ and close to null infinity the
null stress-energy tensor density $\langle T^{\pm}\rangle_{ren}{}_{k_{\pm}k_{\pm}}\varepsilon_{\pm q}$
should not depend on any type of radial parameter of the model, so
that it can be expected on reasonable ground that the condition$\underset{\mathcal{I}^{\pm}}{\int}\langle T^{\pm}\rangle_{ren}{}_{k_{\pm}k_{\pm}}\varepsilon_{\pm q}\geq0$
is actually satisfied. Unfortunately, given that almost half a century
after Hawking's publications on particle creation effects in black
hole physics, there still no solutions to the semiclassical Einstein
equations have been found that describe quantum backreation effects
in dynamical black hole spacetimes, it proves impossible to check
the validity of the assumptions made by explicit calculations.

Having clarified that, one may ultimately use integral relation $(43)$
to set up the semiclassical quasilocal charges

\begin{align}
\mathfrak{Q}_{\pm}= & \underset{r\rightarrow\infty}{\lim}H_{h}^{Boundary}=\\
=\frac{1}{4\pi} & \underset{\mathcal{I}^{\pm}}{\int}\left[N\mathcal{L}_{k_{\pm}}N\Xi_{\pm}+N^{2}[\varkappa_{\pm}\Xi_{\pm}-\sigma'_{\pm ab}\sigma_{\pm}'{}^{ab}+\omega'_{\pm ab}\omega_{\pm}'{}^{ab}-8\pi\mathcal{T}^{\pm}{}_{k_{\pm}k_{\pm}}]\right]\varepsilon_{\pm q},\nonumber 
\end{align}
which take the form

\begin{equation}
\mathfrak{Q}_{\pm}=-\frac{1}{4\pi}\underset{\mathcal{I}^{\pm}}{\int}\left[\vert\mathcal{\mathtt{N}}_{\pm}\vert^{2}+8\pi\mathcal{T}^{\pm}{}_{k_{\pm}k_{\pm}})\right]\varepsilon_{\pm q}
\end{equation}
in Bondi coordinates close to past and future null infinty. Alternatively,
the choices $t_{\pm}^{a}=fk_{\pm}^{a}+V_{\pm}^{a}$ made in the first
section lead to the related semiclassical Brown-York quasilocal charges

\begin{equation}
\mathfrak{Q}_{f}^{\pm}=\frac{1}{4\pi}\underset{\mathcal{I}^{\pm}}{\int}\left[\vert\mathcal{\mathtt{N}}_{\pm}\vert^{2}-\frac{1}{4}D_{a}D_{b}\mathtt{N}_{\pm}^{ab}+8\pi\mathcal{T}^{\pm}{}_{k_{\pm}k_{\pm}}\right]f\varepsilon_{\pm q},
\end{equation}
where again some important details of the derivation are given in
Appendix $A$ of this work.

Given these charges, as further clarified in more detail in the next
section, the anipodal matching conditions $(17)$ can be readily generalized
using the fact that $\mathfrak{Q}_{Y}^{\pm}\equiv Q_{Y}^{\pm}.$ This,
however, prompts the question of how these generalized matching conditions
can be fulfilled in the given physical setting. That question arises
in particular because $\mathcal{H}$ is not modeled as a quasi-stationary
event horizon in the given setting, as is usually the case, but as
a dynamical horizon that changes its causal structure in the course
of its evolution. The latter has an impact on the asymptotic completeness
of the theory and thus on the question of whether information can
be lost in the course of the evaporation process of a black hole.
This will now be explained in more detail in the following, final
subsection of this paper.

\subsection{Dynamical Horizons and Black Hole Evaporation: New Light on Gravitational
Scattering and Information Loss}

The event horizon of a black hole is the future boundary of the spacetime
region from which causal signals can reach $\mathcal{I}^{+}$. The
mere existence of an event horizon thus presupposes knowledge of the
complete Cauchy evolution of spacetime, from the very beginning to
the very end. But this implies that if such a fixed, unchanging hypersurface
were to form in the course of the formation and evaporation process
of a black hole, it would have to exist at all times, even if the
black hole has long since evaporated. The latter, however, is clearly
impossible, since a black hole horizon should obviously exist no longer
than the black hole singularity that created it. Once the singularity
has completely evaporated, at least provided that there is no eternal
black hole remnant at the end of the evaporation process, but the
black hole actually evaporates in its entirety (as Hawking suggested
in his early works on the subject \cite{hawking1974black,hawking1975particle}),
an event horizon cannot actually be part of a realistic physical model
for the formation and evaporation of a black hole. Its extremely global,
teleological geometric character simply does not fit the requirements
of such a model.

So what if it were a dynamical horizon and not an event horizon that
forms during a black hole formation and evaporation process? Would
there be any substantial differences, if a quasilocal description
of the process were given? 

As the arguments set out in \cite{ashtekar2020black,hayward2005disinformation}
clearly show, there would indeed be such substantial differences.
Since the inner boundary of spacetime $(\mathcal{M},g)$ was chosen
to be a dynamical horizon rather than an event horizon, however, the
geometric framework introduced in the first section of this work permits
these arguments of the authors to be followed closely and, at the
same time, to provide a quasilocal description of the process when
$(\mathcal{M},g)$ is assumed to be a nonstationary black hole spacetime.
Given that such a nonstationary black hole spacetime is actually being
considered, the following essential differences arise if the event
horizon of a black hole is replaced by a dynamical horizon:

First, in contrast to the event horizon, the dynamical horizon can
be localized quasilocally and its properties are directly related
to the physical processes at its location. Second, while event horizons
are null hypersurfaces, dynamical horizons can either be spacelike
or timelike or become null; but the latter only if they are isolated
in the sense that no energy flows across them. Third, an event horizon
is only one-way traversable, whereas a dynamical horizon is two-way
traversable. That is, while nothing can escape from the trapped region
enclosed by an event horizon to the 'outside', causal signals can
be sent across a dynamical horizon from 'inside' to 'outside'. Thus,
there is no causal obstruction for the modes that have entered the
trapped region bounded by a dynamical horizon to exit again across
its timelike portion. 

Against this background, the following evaporation scenario can be
envisioned, which is largely based on the assessment of the situation
given in \cite{ashtekar2020black}: As the evaporation process sets
in, the mass of the black hole starts to decrease and the area of
the marginally trapped surfaces of the black hole goes down, thus
violating the law that, classically, the area cannot decrease. This
violation must apparently be caused by some flux of negative energy
across the dynamical horizon into the 'interior black hole region',
which balances the positive energy flux radiated towards infinity.
Heuristically, the mentioned negative energy flux should, as one would
expect, be caused by quantum fluctuations leading to spontaneous (stimulated)
emission, where modes are created in pairs, one of which passes the
timelike portion of the dynamical horizon, while the other goes out
to $\mathcal{I}^{+}$. Since the horizon of the black hole changes
dynamically over time, this microscopic pair creation process has
the effect that no lightlike Killing horizon is formed in the course
of the evolution of the black hole, which could eventually be identified
as its event horizon. 

This scenario, however, can now be compared with the toy model of
a Vaidya null fluid spacetime with monotonically decreasing (nonnegative)
mass discussed in Appendix B of this paper. In doing so, obviously
significant differences arise:

First, according to the considered toy model, there is the period
$v<0$ in which the geometry of the black hole corresponds to that
of an eternal Schwarschild black hole. This local spacetime geometry
\cite{huber2020junction}, when considered on its own right, has a
null Killing horizon at $r=2M_{0}$, which would certainly be identified
as the black hole's event horizon if one were to forget the rest of
the black hole's evolution. In the time period $v>0$, this null horizon
however transitions into a dynamical horizon. The mass of the black
hole continues to decrease until it eventually approaches zero and
the singularity at $r=0$ thus disappears. As a result, the dynamical
horizon at $r(v)=2M(v)$ eventually disappears as well. Spacetime
therefore becomes flat at late times. In a semiclassical context,
however, it cannot be expected at all that a 'local event horizon',
i.e. a local null Killing horizon, is formed during black hole evolution.

Second, there is yet another aspect in which a fully-fledged semiclassical
model should differ from the Vaidya toy model considered, even though
the latter has essentially the same mass dependence at future null
infinity \cite{wald1994quantum}: In the Vaidya solution one has an
everywhere positive energy flux that becomes locally large near the
horizon, while in a semiclassical model the energy flux should be
locally negative near the horizon and locally small everywhere 'outside'
the black hole (until the black hole reaches Planck dimensions).

So, there are obvious differences between the toy model used and a
semiclassical model that accounts for the complete formation and evaporation
of black holes. Still, there are some effects revealed by this toy
model that can be expected to occur in the semiclassical regime of
quantum gravity as well.

In particular, since spacetime should become flat again at late times,
all field modes, including the horizon modes, can be expected to converge
to the Minkowskian field modes, analogous to the Vaidya model, if
the mass of the black hole actually reaches a value of zero, i.e.

\begin{equation}
f_{m}^{+}\rightarrow f_{m}^{(0)+},\text{\ensuremath{g_{m}\rightarrow f_{m}^{(0)+}}}
\end{equation}
as $M\rightarrow0$. As shown in Appendix B of this work, the Vaidya
field modes show exactly this kind of behavior, and 

\begin{equation}
\kappa^{-1}\rightarrow0
\end{equation}
is also found to apply as $M\rightarrow0$. The latter has the effect
that, by exploiting the fact that the operators

\begin{equation}
\tilde{a}_{k}^{+}\equiv\frac{1}{\sqrt{2}}(a_{k}^{+}+b_{k}),\;\tilde{a}_{k}^{+\dagger}\equiv\frac{1}{\sqrt{2}}(a_{k}^{+\dagger}+b_{k}^{\dagger}),
\end{equation}
form again a CCR algebra, the decomposition $(30)$ of $\varphi(x)$
with respect to $\mathcal{I}^{+}\cup\mathcal{H}$ reduces to the form

\begin{equation}
\varphi=\overset{\infty}{\underset{k=1}{\sum}}[f_{k}^{(0)+}\tilde{a}_{k}^{+\dagger}+\bar{f}_{k}^{(0)+}\tilde{a}_{k}^{+}].
\end{equation}
Based on the fact that the horizon modes and the other field modes
take this particular form at late times, this, however, implies that
a decomposition of $\varphi(x)$ can be performed with respect to
purely outgoing field modes. The reason for this is that the surface
gravity $\kappa$ of the black hole is not constant, but changes with
time. This has the consequence that $(50)$ should be fulfilled as
$M\rightarrow0$, so that a purification process sets in, which ensures
that eventually no thermal Hawking radiation can be detected at future
null infinity. 

In the semiclassical case, as already clarified, the situation is
completely different. There, a decomposition of $\varphi(x)$ of the
form $(30)$ can of course be performed, but it does not have the
same meaning as for the proposed Vaidya toy model discussed in Appendix
B of this paper. The reason for this is again that the horizon of
spacetime is not a one-way traversable event horizon, but a two-way
traversable dynamical horizon, through which matter and radiation
can escape to null infinity. So there is no extended time period as
in the Vaidya model where spacetime is quasi-static as the black hole
forms and continuously evaporates with a dynamical horizon that is
never truly isolated, i.e. never a truly non-expanding null hypersurface. 

\begin{figure}
\includegraphics[viewport=-86.4322bp 0bp 400.469bp 547.822bp,scale=0.4]{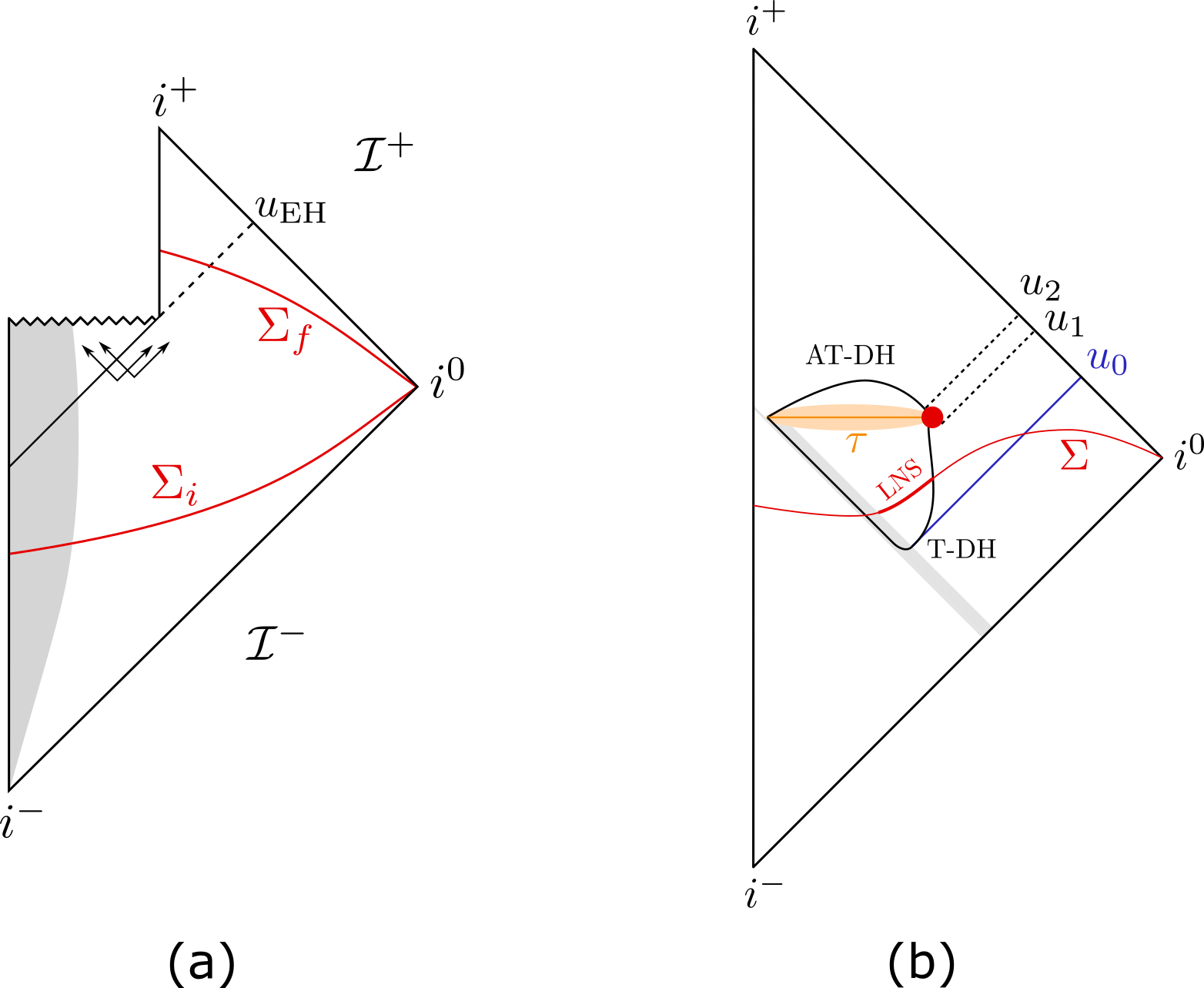}\caption{Graphic $(a)$ to the left shows the commonly used Penrose diagram
to depict black hole evaporation and the geometric structure of the
event horizon throughout the evaporation process, accounting for semiclassical
quantum backreaction effects. Graphic $(b)$ on the right, on the
other hand, shows a possible quantum extension of spacetime as proposed
by Ashtekar in \cite{ashtekar2020black}. Here, the classical singularity
is replaced by a transition hypersurface $\tau$, in the past of which
lies a trapped region bounded in the past by a dynamical trapping
horizon $T-DH$, and in the future of which lies an untrapped region
bounded by a dynamical anti-trapping horizon ($AT-DH$). The graphic
further displays that a purification process occurs in the future
of the transition region $\tau$.}
\end{figure}

Equipped with such a horizon, the black hole evolves significantly
differently than originally pictured by Hawking in \cite{hawking1975particle,hawking1976breakdown},
as incoming radiative modes with negative energy can now continually
affect the black hole singularity as the latter evaporates, so that
there is no reason to expect a thermal Planck spectrum of emitted
Hawking radiation at null infinity. Rather, the occurrence of such
a spectrum can only be expected for a stationary black hole if the
backreaction of the radiation on the singularity of the black hole
is ignored. 

Yet, since in the considered case matter and radiation can escape
through the timelike portion of the black hole horizon, thereby carrying
information, it can be reasonably expected that quasilocal Brown-York
charges of the form $(48)$ can likewise reach future null infinity.
These charges, as has been shown, take the form of generalized quantum
analogues of the supertranslation hairs $(11)$ and superrotation
charges $(16)$ first derived in \cite{flanagan2017conserved,hawking2017superrotation},
which is why these charges should clearly be expected to satisfy the
generalized antipodal matching conditions 
\begin{equation}
\mathfrak{Q}_{f,Y}^{+}=\mathfrak{Q}_{f,Y}^{-}.
\end{equation}
The crucial observation now is that the conditions $(49-52)$ are
principally strong enough to ensure that the latter is indeed the
case, and furthermore to avoid any increase in entropy and thus any
associated loss of quantum information along the way. This is because
the asymptotic form $(52)$ of the scalar field suggests that radiative
modes escaping to future infinity should apparently no longer be thermal
in nature at the end of the evaporation process. Consequently, there
appears to be no information loss in the considered case, since $i)$
the dangerous horizon mode functions take the form of Minkowskian
modes near future null infinity and $ii)$ a Bogoluibov transformation
of the scalar field $(29)$ into one of the form $(52)$ should lead
to a unitary $S$-matrix, since only ingoing and outgoing Minkowskian
field modes are being involved. Hence, it may be expected on reasonable
grounds that everything that has fallen into the black hole must eventually
re-emerge once the black hole is fully evaporated, albeit somewhat
mangled; an expectation that proves consistent with previous results
on the subject both from loop quantum gravity and superstring theory,
which point to the same ultimate outcome of a black hole formation
and evaporation process \cite{ashtekar2005black,ashtekar2005quantum,bousso2002holographic,hooft1996scattering,maldacena2003eternal,mathur2009information,mathur2009exactly,strominger1996microscopic,susskind1997black}.

\section*{Summary and Conclusion}

Using standard techniques from the Brown-York approach to Einstein-Hilbert
gravity, quasilocal charges were derived in the present work that
coincide in the large sphere limit, as was demonstrated, with conserved
the supertranslation hair and superrotation charges introduced by
Hawking, Perry and Strominger in \cite{hawking2017superrotation}.
Based on the results obtained, it was shown that the derived charges
belong to a more general class of quasilocal Brown-York charges, which
was recently derived in \cite{huber2023quasilocal} to elucidate quasilocal
corrections to the Bondi mass loss formula. By incorporating the semiclassical
Einstein equations and resulting quantum backreaction effects, with
a focus placed upon effects generated by massless scalar quantum fields,
a generalization of the deduced soft black hole hair charges was given.
Using these generalized BMS charges, which are sensitive to quantum
backreactions of the spacetime geometry, a general scenario was sketched
in which a black hole completely evaporates after its initial collapse
due to particle creation effects, whereby a genuine event horizon
never forms, but only a slowly evolving dynamical (resp. future trapping)
horizon. Based on the geometric model of a Vaidya-type sandwich spacetime
presented in Appendix B, it was then made clear that, according to
the scenario described, it should be possible for a flux of quasilocal
mass and/or radiant energy to escape through the dynamical horizon
of an evaporating black hole to future null infinity. The quasilocal
charges (arguably) responsible for this energy transfer were identified
in this context as the quantum BMS charges derived, which generalize
those of Hawking, Perry and Strominger. Ultimately, by imposing conditions
on the mode functions of the scalar field, again derived with respect
to the Vaidya toy model used, it was made clear that particle creation
effects in nonstationary black hole spacetimes with locally defined
dynamical horizons arguably cannot cause information loss at the quantum
level.

That said, it may be noted that the the only evaporation scenario
covered by this article is, as already emphasized, one in which the
black hole fully evaporates due to the emission of Hawking radiation;
thereby abandoning the possibility of the formation of a naked singularity
\cite{goswami2014collapsing,ziaie2011naked} or a black hole remnant
in the Planck scale regime (see here, for example, \cite{chen2015black}
and references therein for information on the subject). Yet, it may
also be noted that the results obtained, while showing that the singularity
is naked in phases in which the black hole horizon is dynamical and
thus two-way traversable, do not rule out the possibility of the formation
of a black hole remnant. The main motivation for nonetheless discarding
this possibility and considering only the full evaporation scenario
treated above is rooted in the main objective of the present work,
which is to show that the results of \cite{hawking1974black,hawking1975particle,hawking1976breakdown,hawking1982unpredictability}
and \cite{hawking2016soft,hawking2017superrotation} prove to be consistent
and that, therefore, Hawking's early ideas on the subject can be reconciled
with his later ideas.

The dynamical black hole spacetime considered for this purpose was
modeled at the semiclassical level, which is why no singularity resolution
was taken into account. This is in sharp contrast to other contemporary
quantum gravity approaches to the information paradox, such as loop
quantum gravity and superstring theory, where there are fundamental
length scales that rule out the existence of singularities in black
holes from the outset. In such more general theories of gravity, which
surpass the semiclassical regime, singularity resolution would certainly
have to be taken into account. Still, it can be expected on reasonable
grounds that some type of similar evaporation scenario could occur
as well in the theories mentioned (albeit presumably in a modified
form). However, only future work in these fields will show whether
the consideration of purification scenarios, as outlined in the present
work, and related arguments against a loss of quantum information
in evaporating black hole spacetimes will or will not actually prove
useful in genuine theories of quantum gravity.

As for now, it looks like the adopted geometric framework is potentially
reconcilable with the concept of holographic screens and other aspects
of the theory of holography \cite{bousso2017dynamics}. After all,
both approaches are based on a similar null geometric foundation in
the classical regime (though the concept of a dynamical horizon considered
in this work generally differs from that of a holographic screen),
and there are no strong reasons to believe that they should not prove
compatible in the semiclassical regime as well. Admittedly, however,
this point merits a closer examination of the subject in the future.

Ultimately, the question arises as to the experimental verification
of the ideas presented. More specifically, since there are a number
of approaches that suggest (or at least strive for) verification of
Hawking's results in black hole physics, e.g. through gravitational
wave data or imaging of black holes or even through reallife laboratory
experiments \cite{barcelo2003towards,belgiorno2010hawking,munoz2019observation,schutzhold2005hawking,steinhauer2016observation,weinfurtner2013classical},
the question naturally arises whether purification processes and quantum
information preserving effects, as presented in this paper, could
also be observed or measured in the laboratory. The expectation here
might be that methods similar to those likely to be used to confirm
the existence of Hawking radiation could also be used to confirm evidence
for the radiation purification mechanism presented in this paper.
But this is, of course, pure speculation. Still, such confirmation
would clearly be desirable, not least because it is generally expected
that the results obtained to date in black hole thermodynamics are
fundamental in the sense that they should also remain valid within
the deep Planckian regime of quantum gravity. Seen in this light,
any experimental evidence, however small, could ultimately shed light
not only on a long-standing paradox that has attracted the greatest
interest of researchers ever since its formulation almost half a century
ago, but also on a theory of quantum gravity that has yet to be explored
in full detail. 

\section*{Appendix A}

The purpose of this part of the manuscript is to further clarify how
the Brown-York charges constructed in the main part of the work lead
to the soft supertranslation hairs and superrotation charges constructed
in \cite{flanagan2017conserved,hawking2017superrotation}. Considering
to this end the geometric framework laid out in these works, the construction
is to be carried out in retarded coordinates $(u,r,x^{A})$ near future
null infinity $\mathcal{I}^{+}$ and advanced coordinates $(v,r,x^{A})$
near future past infinity $\mathcal{I}^{-}$ in the Bondi gauge. In
this gauge, $g_{rr}=g_{rA}=0$ and $g_{AB}=r^{2}q_{AB}$ with $\partial_{u}q=\partial_{r}q=0$
resp. $\partial_{v}q=\partial_{r}q=0$ applies, where $q$ denotes
the determinant of the induced two-metric $q_{AB}$. As a consequence,
the spacetime metric takes the form
\begin{equation}
ds^{2}=-\frac{U}{r}e^{-2\beta_{+}}du^{2}-2e^{2\beta_{+}}dudr+r^{2}q_{AB}^{+}(dx^{A}-U^{A}du)(dx^{A}-U^{A}du)
\end{equation}
in retarded coordinates, whereas in advanced coordinates it reads

\begin{equation}
ds^{2}=-\frac{V}{r}e^{-2\beta_{-}}dv^{2}+2e^{2\beta_{-}}dvdr+r^{2}q_{AB}^{-}(dx^{A}-V^{A}dv)(dx^{A}-V^{A}dv).
\end{equation}
This form of the metric (resp. that of the associated inverse metric)
then allows one to define the null vector fields $l^{a}=-\partial_{r}^{a}$,
$k_{+}^{a}=e^{-2\beta}(\partial_{u}^{a}+\frac{U}{2r}\partial_{r}^{a}+U^{A}\partial_{A}^{a})$
and $k_{-}^{a}=e^{-2\beta}(\partial_{v}^{a}+\frac{V}{2r}\partial_{r}^{a}+V^{A}\partial_{A}^{a})$.
Using additionally the expansions $\frac{U}{r}e^{-2\beta_{+}}=\frac{V}{r}e^{-2\beta_{-}}=1-\frac{2m}{r}+\mathscr{O}(r^{-2})$,
$e^{2\beta_{\pm}}=1-\frac{C_{AB}^{\pm}C_{\pm}^{AB}}{32r^{2}}+\mathscr{O}(r^{-3})$,
$U^{A}=u_{A}+\mathscr{O}(r^{-4})$ with $u_{A}:=-\frac{1}{2}D^{B}C_{+\:B}^{A}-\frac{1}{r^{3}}[\frac{2}{3}\mathcal{\mathtt{N}}_{A}+\frac{2u}{3}D_{A}m-\frac{1}{16}D_{A}(C_{CD}^{+}C_{+}^{CD})]$,
$v_{A}:=-\frac{1}{2}D^{B}C_{-\:B}^{A}-\frac{1}{r^{3}}[\frac{2}{3}\mathcal{\mathtt{N}}_{A}+\frac{2v}{3}D_{A}m-\frac{1}{16}D_{A}(C_{CD}^{-}C_{-}^{CD})]$
and $q_{AB}^{\pm}=(1+\frac{1}{4r^{2}}C_{CD}^{\pm}C_{\pm}^{CD})\gamma_{AB}+\frac{1}{r}C_{AB}^{\pm}+\mathscr{O}(r^{-3})$,
where $\mathcal{\mathtt{N}}_{A}$ is the angular momentum aspect and
$\gamma_{AB}$ is a the induced metric on the sphere, the metric reduces
to the form

\begin{equation}
ds^{2}=-(1-\frac{2m}{r})du^{2}-2dudr+u_{A}dudx^{A}+r^{2}q_{AB}^{+}dx^{A}dx^{B}
\end{equation}
resp.

\begin{equation}
ds^{2}=-(1-\frac{2m}{r})dv^{2}+2dvdr+v_{A}dvdx^{A}+r^{2}q_{AB}^{-}dx^{A}dx^{B}
\end{equation}
in the introduced coordinates. To identify the lapse function $N$
of the geometry, it proves convenient to introduce new coordinates
$u=t-r$ and $v=t+r$. In these coordinates $(51)$ and $(52)$ can
be rewritten in the form 
\begin{align}
ds^{2} & =-(1-\frac{2m}{r})dt^{2}-\frac{4m}{r}dtdr+(1+\frac{2m}{r})dr^{2}+u_{A}(dt-dr)dx^{A}\\
 & +r^{2}q_{AB}^{+}dx^{A}dx^{B}\nonumber 
\end{align}
resp.

\begin{align}
ds^{2} & =-(1-\frac{2m}{r})dt^{2}+\frac{4m}{r}dtdr+(1+\frac{2m}{r})dr^{2}+v_{A}(dt+dr)dx^{A}\\
 & +r^{2}q_{AB}^{-}dx^{A}dx^{B},\nonumber 
\end{align}
thereby allowing one to identify $N(r)=\sqrt{1-\frac{2m}{r}}$ as
the lapse function to the geometry. That said, a possible choice for
the null vector fields of the geometry in the given coordinate setting
is $l^{a}=-\partial_{r}^{a}$, $k_{+}^{a}=(\partial_{u}^{a}-\frac{N}{2}\partial_{r}^{a})$
and $k_{-}^{a}=(\partial_{v}^{a}+\frac{N}{2}\partial_{r}^{a})$. The
latter, however, implies that $\partial_{u}^{a}=k_{+}^{a}-\frac{N}{2}l^{a}$
and $\partial_{v}^{a}=k_{+}^{a}+\frac{N}{2}l^{a}$. 

Given all that, the stress-energy tensor of spacetime shall be assumed
to satisfy the following fall-off conditions as $r\rightarrow\infty$:
$T_{uu}(u,r,x^{A})=\frac{\mathfrak{T}{}_{uu}(u,x^{A})}{r^{2}}+\mathscr{O}(r^{-3})$,
$T_{uA}(u,r,x^{A})=\frac{\mathfrak{T}_{uA}(u,x^{A})}{r^{2}}+\mathscr{O}(r^{-3})$,
$T_{AB}(u,r,x^{A})=\frac{\mathfrak{T}(u,x^{A})\gamma_{AB}}{r}+\mathscr{O}(r^{-2})$,
$T_{ur}(u,r,x^{A})=\mathscr{O}(r^{-4})$, $T_{rr}(u,r,x^{A})=\mathscr{O}(r^{-4})$,
$T_{rA}(u,r,x^{A})=\mathscr{O}(r^{-3})$ resp. $T_{vv}(v,r,x^{A})=\frac{\mathfrak{T}_{vv}(v,x^{A})}{r^{2}}+\mathscr{O}(r^{-3})$,
$T_{vA}(v,r,x^{A})=\frac{\mathfrak{T}_{vA}(v,x^{A})}{r^{2}}+\mathscr{O}(r^{-3})$,
$T_{AB}(v,r,x^{A})=\frac{\mathfrak{T}(v,x^{A})\gamma_{AB}}{r}+\mathscr{O}(r^{-2})$,
$T_{vr}(v,r,x^{A})=\mathscr{O}(r^{-4})$, $T_{rr}(v,r,x^{A})=\mathscr{O}(r^{-4})$,
$T_{rA}(v,r,x^{A})=\mathscr{O}(r^{-3})$. These conditions yield
\begin{align}
T_{kk}^{+}= & \underset{r\rightarrow\infty}{\lim}r^{2}T_{ab}k_{+}^{a}k_{+}^{b}=\underset{r\rightarrow\infty}{\lim}r^{2}T_{uu}=\mathfrak{T}_{uu},\\
T_{kk}^{-}= & \underset{r\rightarrow\infty}{\lim}r^{2}T_{ab}k_{-}^{a}k_{-}^{b}=\underset{r\rightarrow\infty}{\lim}r^{2}T_{vv}=\mathfrak{T}_{vv}.\nonumber 
\end{align}
Moreover, introducing next the past and future news tensors $\mathcal{\mathtt{N}}_{AB}^{+}=\partial_{u}C_{AB}^{+}$
and $\mathcal{\mathtt{N}}_{AB}^{-}=\partial_{v}C_{AB}^{-}$, where
$C_{AB}^{\pm}=-2D_{A}D_{B}C\vert_{\mathcal{I}_{\mp}^{\pm}}+\gamma_{AB}D^{2}C\vert_{\mathcal{I}_{\mp}^{\pm}}$
with $C^{+}=C^{+}(u,x^{A})$ and $C^{-}=C^{-}(v,x^{A})$ shall apply
by definition, let it be be required - along the same lines as in
\cite{hawking2017superrotation} - that near the past and future boundaries
of $\mathcal{I}^{\pm}$ and $\mathcal{I}_{\mp}^{\pm}$, the news falls
off faster than $\frac{1}{\vert u\vert}$ resp. $\frac{1}{\vert v\vert}$
and that the angular momentum aspect $\mathcal{\mathtt{N}}_{A}$ approaches
a finite one-form on $\mathbb{S}_{2}$. This then makes it clear that
the bulk is not required to be nearly flat far from the boundary,
but can have any geometric structure whatsoever.

Using the above, it then becomes clear that the quasilocal charges
depicted in $(10)$ take the form
\begin{align}
Q_{+} & =\frac{1}{4\pi}\underset{\mathcal{I}^{+}}{\int}\left[\frac{1}{8}\mathcal{\mathtt{N}}_{AB}^{+}\mathcal{\mathtt{N}}_{+}^{AB}+8\pi\mathfrak{T}_{uu}\right]\varepsilon_{+\gamma},\\
Q_{-} & =\frac{1}{4\pi}\underset{\mathcal{I}^{-}}{\int}\left[\frac{1}{8}\mathcal{\mathtt{N}}_{AB}^{-}\mathcal{\mathtt{N}}_{-}^{AB}+8\pi\mathfrak{T}_{vv}\right]\varepsilon_{-\gamma}.\nonumber 
\end{align}
thus leading one to conclude that the expressions obtained coincide
- up to a total divergence term equal to zero - with the families
of charges defined in\footnote{To be precise, the charges occur in the special case $f=1$.}
\cite{hawking2017superrotation}. Moreover, taking relations $(14-16)$
into account, one obtains the further results

\begin{align}
J_{\mathcal{I}_{-}^{+}}^{\phi}=\frac{1}{4\pi}\underset{\mathcal{I}_{-}^{+}}{\int}\Omega_{A}\phi^{A}\omega_{\gamma} & =\frac{1}{8\pi}\underset{\mathcal{I}_{-}^{+}}{\int}\mathcal{\mathtt{N}}_{A}Y^{A}\omega_{\gamma}=Q_{Y}^{+},\\
J_{\mathcal{I}_{+}^{-}}^{\phi}=\frac{1}{4\pi}\underset{\mathcal{I}_{+}^{-}}{\int}\Omega_{A}\phi^{A}\omega_{\gamma} & =\frac{1}{8\pi}\underset{\mathcal{I}_{+}^{-}}{\int}\mathcal{\mathtt{N}}_{A}Y^{A}\omega_{\gamma}=Q_{Y}^{-},\nonumber 
\end{align}
 provided that $\mathcal{\mathtt{N}}^{A}$ denotes the angular momentum
aspect and $Y^{A}$ is a conformal Killing vector field on the two-sphere
given by $Y^{A}=D^{A}\chi+\varepsilon^{AB}D_{B}\kappa$ with $\chi$
and $\kappa$ representing solutions of the differential relation
$(D^{2}+2)\chi=(D^{2}+2)\kappa=0$ for $D^{2}:=D_{A}D^{A}.$

Considering now as in \cite{hawking2017superrotation} a supertranslation
generating diffeomorphism vector fields of the form $\zeta_{+}^{a}=f\partial_{u}^{a}-\frac{1}{r}D^{A}f\partial_{A}^{a}+\frac{1}{2}D^{2}f\partial_{r}^{a}$
with $\partial_{u}f=\partial_{r}f=0$ and $\zeta_{-}^{a}=f\partial_{v}^{a}-\frac{1}{r}D^{A}f\partial_{A}^{a}+\frac{1}{2}D^{2}f\partial_{r}^{a}$
with $\partial_{v}f=\partial_{r}f=0$, the same line of reasoning
that led to the form of $(61)$ can be used to conclude that Brown-York
charges $(11)$ reduce to the form 
\begin{align}
Q_{f}^{+} & =\frac{1}{4\pi}\underset{\mathcal{I}^{+}}{\int}\left[\frac{1}{8}\mathtt{N}_{AB}^{+}\mathtt{N}_{+}^{AB}-\frac{1}{4}D_{A}D_{B}\mathtt{N}_{+}^{AB}+8\pi\mathfrak{T}_{uu}\right]f\varepsilon_{+\gamma},\\
Q_{f}^{-} & =\frac{1}{4\pi}\underset{\mathcal{I}^{-}}{\int}\left[\frac{1}{8}\mathtt{N}_{AB}^{-}\mathtt{N}_{-}^{AB}-\frac{1}{4}D_{A}D_{B}\mathtt{N}_{-}^{AB}+8\pi\mathfrak{T}_{vv}\right]f\varepsilon_{-\gamma},\nonumber 
\end{align}
if $t^{a}\equiv\zeta_{\pm}^{a}$ is chosen as the time evolution vector
field of spacetime and the boundary of the latter is shifted to infinity
in the large sphere limit, where it has been used that $\vert\mathcal{\mathtt{N}}_{\pm}\vert^{2}=\frac{1}{8}\mathtt{N}_{AB}^{\pm}\mathtt{N}_{\pm}^{AB}$.
This shows that the conserved supertranslation hair and superrotation
charge expressions arise naturally in the given coordinate setting
from the quasilocal Brown-York charges defined in section one and
two of this manuscript. 

That said, the antipodal matching conditions $(17)$ are found to
be equivalent to the Lorentz and CPT invariant matching conditions
\begin{equation}
C\vert_{\mathcal{I}_{-}^{+}}=C\vert_{\mathcal{I}_{+}^{-}},\:m\vert_{\mathcal{I}_{-}^{+}}=m\vert_{\mathcal{I}_{+}^{-}},\:N^{A}\vert_{\mathcal{I}_{-}^{+}}=N^{A}\vert_{\mathcal{I}_{+}^{-}}
\end{equation}
formulated in \cite{hawking2017superrotation}, thereby proving consistency
with the Brown-York approach adopted in this work.

Eventually, given the transition $g_{ab}\rightarrow g_{ab}+L_{\zeta}g_{ab}$,
where the variations $L_{\zeta}g_{ab}$ around the background metric
$g_{ab}$ are typically assumed to obey the linearized Einstein equations,
it was shown in \cite{hawking2017superrotation} that one can now
use the foregoing to $i)$ 'implant' supertranslation hairs in black
hole spacetimes and $ii)$ construct horizon charges whose explicit
form depends on the choice of coordinates used to extend the supertranslations
in from the boundary to the horizon. 

Such charges are constructed explicitly in \cite{hawking2017superrotation}.
It is worth noting that the above results show that the construction
of such horizon charges or supertranslation hair in black hole spacetimes
is, in principle, compatible with the Brown-York framework used in
the present work, even though said horizon charges have not been considered
in the main body of the manuscript.

Instead quasilocal charge resulting from the semiclassical Einstein
equations $(37)$ were calculated, where the latter may be rewritten
in the form 
\begin{equation}
E_{\;b}^{a}=G_{\;b}^{a}-8\pi\mathcal{T}_{\;b}^{a}=0.
\end{equation}
To perform a semiclassical extension of the standard Bondi-Sachs formalism
(see here, for instance, \cite{madler2016bondi} and references therein
for an introduction to the subject), it shall be assumed in the following
that a renormalization scheme is used according to which the renormalized
expectation value of the stress-energy tensor (in combination with
all resulting higher-order curvature corrections) is covariantly conserved.
Under this assumption, the contracted Bianchi identity
\begin{equation}
\nabla_{a}E_{\;b}^{a}=0
\end{equation}
is satisfied in the same way as in the classic Bondi-Sachs approach.
In consequence, the hypersurface and evolution equations
\begin{equation}
E_{\;a}^{u}=0,\;E_{AB}-\frac{1}{2}q_{AB}E=0,
\end{equation}
resp.
\begin{equation}
E_{\;a}^{v}=0,\;E_{AB}-\frac{1}{2}q_{AB}E=0,
\end{equation}
can be designated close to future and past null infinity, where $E=q^{CD}E_{CD}$
with $C,D=2,3$ shall apply by definition. For the purpose of integrating
these main equations, in principle, the same steps as in the standard
classical Bondi-Sachs formalism can be taken. In particular, semiclassical
metric data similar to those in $(54)$ and $(55)$ can be prescribed
as a basis for the integration of $(67)$ and $(68)$, which in turn
leads to line elements of the form $(56)$ and $(57)$ if the same
fall-off conditions as at the beginning of this Appendix are imposed.
Eventually, if the same fall-off conditions are invoked for the renormalized
expectation value of the stress-energy tensor operator as in the above
case for the classical energy-momentum tensor, i.e., if $T_{ab}(x)$
and $\mathcal{T}_{ab}(x)$ exhibit the same asymptotic fall-off behavior,
the formalism used yields semiclassical analogues of the quasilocal
Brown-York charges $(10)$ and $(11)$, the explicit form of which
is depicted in $(47)$ and $(48)$.

\section*{Appendix B}

As it proves to be too difficult to model a black hole formation and
evaporation process with semiclassical methods, an obvious alternative
is to consider a manageable spherically symmetric model that covers
at least some essential properties of the geometry of an evaporating
black hole. Such a model is given by the Vaidya geometry
\begin{equation}
ds^{2}=-(1-\frac{2M}{r})dv^{2}+2dvdr+r^{2}d\Omega^{2},
\end{equation}
which describes the gravitational field of a null fluid with time-dependent
mass $M(v)$ and a stress-energy tensor of the form
\begin{equation}
T_{\;b}^{a}=\frac{\dot{M}}{4\pi r}k^{a}k_{b}.
\end{equation}
To describe the evaporation part of the evolution of the black hole,
it makes sense to consider the mass function 
\begin{equation}
M(v)=\begin{cases}
\overset{M_{0}}{\underset{0}{M_{0}[1-\frac{v}{v_{0}}]^{\frac{1}{3}}}} & \overset{v<0}{\underset{v>v_{0}}{v\in[0,v_{0}]}}\end{cases}
\end{equation}
with $v_{0}=\frac{3c}{M_{0}^{3}}$. Accordingly, the evaporation process
take places in the time interval $[0,v_{0}]$, whereby, depending
on the concrete value of the constant $c$, it may be assumed that
the geometry of the black hole evolves slowly. The particular choice
made for the mass function in $(71)$ give rises to the evolution
equation 
\begin{equation}
\frac{dM}{dv}=-\frac{c}{M^{2}}
\end{equation}
in the time period $[0,v_{0}]$, in full agreement with Stefan-Boltzmann's
law. The focus is thus solely placed on the evaporation process, whereas
the formation process of the black hole is completely disregarded.
The black hole has simply always existed in the period $v<0$ with
a distributional stress-energy tensor of the form \cite{balasin1993energy,balasin1994distributional}
\begin{equation}
T_{\;b}^{a}=-M_{0}\delta^{(3)}(x)v^{a}v_{b},
\end{equation}
where $v^{a}$ is a timelike vector field 'outside' the $r=2M_{0}$-region.

As can be seen, the geometry has a Kodama vector field $\xi^{a}=\partial_{v}^{a}$
that constitutes a genuine Killing vector field at early and late
times \cite{kodama1980conserved}, where the latter can be identified
as a hidden Killing field in the sense of \cite{huber2021hidden}.
As may further be noted, the dynamical horizon of the geometry lies
at $r_{\mathcal{H}}(v)=2M(v)$. Since the black hole evaporates very
slowly, it can be concluded that $M(v)\approx M_{0}$ at the beginning
of the evaporation period, thereby implying that the dynamical horizon
$\mathcal{H}$ behaves in that time period almost like a quasi-static
black hole horizon with associated constant radial parameter $r_{0}=2M_{0}$,
which can actually be identified as an isolated horizon in the sense
of Ashtekar et al. \cite{ashtekar2000generic,ashtekar1999isolated,ashtekar2004isolated}.
Still, this horizon is not a genuine event horizon. It is merely a
local event horizon for the local black hole spacetime, which exists
in the time span $v\in(-\infty,0]$; for further details on local
spacetimes, see \cite{huber2020junction}. 

In any case, given the above geometric setting, let the wave equation
$(28)$ be considered next using the identity $\boxempty=\frac{1}{\sqrt{-g}}\partial_{a}(\sqrt{-g}g^{ab}\partial_{b})$.
After separating the variables here as

\begin{equation}
\varphi_{\omega lm}(v,r,\theta,\phi)=\frac{1}{r}\rho(v,r;\omega)Y_{lm}(\theta,\phi),
\end{equation}
one finds that the equation splits up in two parts: a standard angular
part, which can straightforwardly be solved and thus will not be of
much interest in the following, and a radial part
\begin{equation}
(1-\frac{2M}{r})\frac{\partial^{2}\rho}{\partial r^{2}}+2\frac{\partial^{2}\rho}{\partial r\partial v}+\frac{2M}{r^{2}}\frac{\partial\rho}{\partial r}-[\frac{2M}{r^{3}}+\frac{l(l+1)}{r^{2}}]\rho=0,
\end{equation}
which will be the main focus of attention in what follows. Following
closely the arguments used in \cite{jianyang1994hawking,zheng1992new},
this radial part can be rewritten by introducing the tortoise coordinate
$r_{*}=r+\frac{1}{2\kappa}\ln\vert\frac{r-r_{\mathcal{H}}}{r_{\mathcal{H}}}\vert$
in the form

\begin{equation}
A\frac{\partial^{2}\rho}{\partial r_{*}^{2}}+2\frac{\partial^{2}\rho}{\partial r_{*}\partial v}+B\frac{\partial\rho}{\partial r_{*}}-C[\frac{2M}{r^{3}}+\frac{l(l+1)}{r^{2}}]\rho=0,
\end{equation}
provided that the definitions $A(v,r)=\frac{(2\kappa(r-r_{\mathcal{H}})+1)(r-2M)-2r\dot{r}_{\mathcal{H}}}{2\kappa(r-r_{\mathcal{H}})}$,
$B(v,r)=\frac{2r\dot{r}_{\mathcal{H}}-r+2M}{r(r-r_{\mathcal{H}})(2\kappa(r-r_{\mathcal{H}})+1)}+\frac{2M}{r^{2}}$
and $C(v,r)=\frac{2\kappa(r-r_{\mathcal{H}})}{2\kappa(r-r_{\mathcal{H}})+1}$
are used in the present context. Supposing that $\underset{r\rightarrow r_{\mathcal{H}}}{\lim}A(v,r)\rightarrow1$,
one then finds

\begin{equation}
\underset{r\rightarrow r_{\mathcal{H}}}{\lim}(r-2M+2r\dot{r}_{\mathcal{H}})\rightarrow0,
\end{equation}
whence
\begin{equation}
r_{\mathcal{H}}=\frac{2M}{1-2\dot{r}_{\mathcal{H}}},\;\kappa=\frac{1-2\dot{r}_{\mathcal{H}}}{4M}.
\end{equation}
For small $\dot{r}_{\mathcal{H}}$ (and hence $\dot{M}$), it can
thus be concluded that
\begin{equation}
r_{\mathcal{H}}\approx2M(1+4\dot{M}),\;\kappa\approx\frac{1-4\dot{M}}{4M}
\end{equation}
in agreement with the results of Babinot \cite{balbinot1986back}
and York \cite{york1984quantum}. At the very beginning of the evaporation
process, one may safely assume that $\dot{M}\approx0$, which, however,
allows one to conclude that

\begin{equation}
r_{\mathcal{H}}\approx2M,\;\kappa_{0}\approx\frac{1}{4M}
\end{equation}
applies in this particular period of time. As a result, equation $(76)$
reduces to the form

\begin{equation}
\frac{\partial^{2}\rho}{\partial r_{*}^{2}}+2\frac{\partial^{2}\rho}{\partial r_{*}\partial v}+2\kappa_{0}\frac{\partial\rho}{\partial r_{*}}=0,
\end{equation}
as follows from the fact that $\underset{r\rightarrow r_{\mathcal{H}}}{\lim}A(r,v)\rightarrow1$,
$\underset{r\rightarrow r_{\mathcal{H}}}{\lim}B(r,v)\rightarrow2\kappa_{0}(v)$
and $\underset{r\rightarrow r_{\mathcal{H}}}{\lim}C(r,v)\rightarrow0$. 

The differential equation thus obtained can be solved. It has the
two independent solutions
\begin{equation}
\rho_{\omega}^{in}=e^{-i\omega v+\mathcal{K}_{0}},\;\rho_{\omega}^{out}=e^{-i\omega(v-2r_{*})+\mathcal{K}_{0}},
\end{equation}
considering that $\mathcal{K}_{0}(v):=\underset{0}{\overset{v}{\int}}\kappa_{0}(v_{*})dv_{*}=\frac{3v_{0}}{2M_{0}}M^{2}(v)+C$
is defined with respect to an integration constant $C$, which for
the sake of simplicity shall be set to zero in the given context. 

The form of these mode solutions clearly shows why it makes sense
to consider $\kappa_{0}(v)$ instead of $\kappa(v)$ in $(72)$ and
to use $\mathcal{K}_{0}(v)$ instead of $\mathcal{K}(v):=\underset{0}{\overset{v}{\int}}\kappa(v_{*})dv_{*}=-\frac{3v_{0}}{2M_{0}}M^{2}(v)-\ln M(v)+C$
for the definition of the mode functions $(73)$: On the one hand,
one has $e^{\mathcal{K}_{0}(v)}\vert_{v>v_{0}}=1$ due to the fact
that $\mathcal{K}_{0}(v)\vert_{v<v_{0}}=0$ applies by definition,
whereas on the other hand one finds after the evaporation period $e^{\mathcal{K}(v)}\vert_{v>v_{0}}=\frac{1}{M(v)}e^{-\frac{3v_{0}}{2M_{0}}M^{2}(v)}\vert_{v>v_{0}}\rightarrow\infty$.
Therefore, it clearly makes more sense to set up relation $(82)$
in the given form; where the latter is closely related to, but yet
not identical with one previously obtained in \cite{liu2012tortoise}
for a different radial parameter $r^{*}=\frac{1}{2\kappa}\ln\vert\frac{r-r_{\mathcal{H}}}{r_{\mathcal{H}}}\vert$
and different types of functions $A(r,v)$, $B(r,v)$ and $C(r,v)$
in $(76)$.

As may be noted, the part $e^{-i\omega(v-2r_{*})}=e^{-i\omega(v-2r)}(r-r_{\mathcal{H}})^{\frac{i\omega}{\kappa_{0}}}$
is not analytic at the horizon $r_{\mathcal{H}}$. Following \cite{zheng1992new},
however, the mentioned part can be analytically continued from $r>r_{\mathcal{H}}$
to $r\leq r_{\mathcal{H}}$ such that
\begin{align}
r-r_{\mathcal{H}} & \rightarrow\vert r-r_{\mathcal{H}}\vert e^{-i\pi}=(r-r_{\mathcal{H}})e^{-i\pi},\\
\rho_{\omega}^{out} & \rightarrow\tilde{\rho}_{\omega}^{out}=e^{\frac{\pi\omega}{\kappa_{0}}}\rho_{\omega}^{out}.\nonumber 
\end{align}
Thus, the relative scattering probability of the outgoing wave at
the horizon is
\begin{equation}
\vert\frac{\rho_{\omega}^{out}}{\tilde{\rho}_{\omega}^{out}}\vert^{2}=e^{-\frac{2\pi\omega}{\kappa_{0}}},
\end{equation}
thereby implying that the spectrum of the Hawking radiation from the
horizon is 
\begin{equation}
N_{\omega}=\frac{1}{e^{\frac{2\pi\omega_{k}}{\kappa_{0}}}-1}.
\end{equation}
From this, in turn, it can be concluded that the black hole emits
thermal Hawking radiation of temperature 
\begin{equation}
T=\frac{\kappa_{0}}{2\pi}.
\end{equation}
The field modes $(83)$ and $(84)$ can therefore be used to perform
a splitting of the form $(30)$ of $\varphi_{\omega lm}(v,r,\theta,\phi)$
(possibly after smearing them out with test functions with compact
support at the dynamical horizon $\mathcal{H}$ and at null infinity).
Still, this splitting cannot simply be regarded as a conventional
splitting into future null infinity field modes and horizon modes.
To see why, one may recall that in the local flat Minkwoskian part
of the spacetime occurring for $v>v_{0}$ relations $(75)$ and $(76)$
reduce to the form
\begin{equation}
\frac{\partial^{2}\rho}{\partial r^{2}}+2\frac{\partial^{2}\rho}{\partial r\partial v}=0,
\end{equation}
thereby giving rise to two independent solutions of the form

\begin{equation}
\rho_{0\omega}^{in}=e^{-i\omega v},\;\rho_{0\omega}^{out}=e^{-i\omega(v-2r)}.
\end{equation}
In the Schwarzschild case occurring for $v<0$, on the other hand,
differential relation $(76)$ becomes 
\begin{equation}
\frac{\partial^{2}\rho}{\partial r_{0*}^{2}}+2\frac{\partial^{2}\rho}{\partial r_{0*}\partial v}+2\kappa_{0}\frac{\partial\rho}{\partial r_{0*}}=0,
\end{equation}
for small variations of the mass parameter (i.e. for $\dot{M}(v)\approx0$
and $M(v)\approx M_{0}\approx const.$), where $r_{0*}=r+2M_{0}\ln\vert\frac{r-2M_{0}}{2M_{0}}\vert$
applies by definition. The solutions of this equation read

\begin{equation}
\rho_{\omega}^{in}=e^{-i\omega v+\mathcal{K}_{0}},\;\rho_{\omega}^{out}=e^{-i\omega(v-2r_{0*})+\mathcal{K}_{0}},\;\tilde{\rho}_{\omega}^{out}=e^{\frac{\pi\omega}{\kappa_{0}}}\rho_{\omega}^{out},
\end{equation}
and thus coincide exactly with $(83)$ and $(84)$ for $M(v)\approx M_{0}$.
In this case, the first law of black hole mechanics \cite{bardeen1973four}
dictates that $\kappa_{0}=const.$ applies, so that it can be concluded
that $\mathcal{K}_{0}(v)=\kappa_{0}v$ applies in the time span $v<0$.
This changes for $v\in[0,\infty)$; a period in which it can be observed
that $\kappa_{0}^{-1}\rightarrow0,\;\mathcal{K}_{0}\rightarrow0$
as $M(v)\rightarrow0$. Consequently, however, it can be concluded
that the dynamical mode functions $(83)$ and $(84)$ at late times
take the form
\begin{equation}
\rho_{\omega}^{in}\rightarrow\rho_{0\omega}^{in},\;\rho_{\omega}^{out}\rightarrow\rho_{0\omega}^{out},\;\tilde{\rho}_{\omega}^{out}\rightarrow\rho_{0\omega}^{out}
\end{equation}
and thus coincide with the Minkowskian mode functions $(88)$ for
$v\in[v_{0},\infty)$. As a result, the radiation escaping to future
null infinity appears no longer to be thermal in nature at the end
of the evaporation process. It is therefore not expected, according
to the proposed toy model, that a thermal spectrum of the form $(35)$
or $(85)$ occurs at late epochs of the evolution of spacetime. 

That said, let it ultimately be noted that comparable results can
also be derived for a supertranslated Vaidya spacetime, as considered
in particular in \cite{chu2018soft}, if a mass parameter of the form
$(71)$ is considered. Despite the fact that the expression for the
surface gravity takes a different form in said case (see here equation
$3.12$ on page $10$), still there is full confirmation of the results
obtained in the present work. Thus, once again, there is good agreement
with the previously known literature on the subject.
\begin{description}
\item [{Acknowledgements:}]~
\end{description}
I want to thank Felix Wilkens for his support in preparing the images
depicted in Figures $1$ and $2$ of the paper. Moreover, I want to
acknowledge the UAS Technikum Wien Library for financial support through
its Open Access Funding Program.

\bibliographystyle{plain}
\addcontentsline{toc}{section}{\refname}\bibliography{2C__Arbeiten_litcha}

\end{document}